\documentclass[longauth]{aaEC}

\usepackage{graphicx}
\usepackage{natbib}
\usepackage{scalerel}
\usepackage{comment}

\usepackage[table]{xcolor}
\usepackage[para]{threeparttable}

\usepackage{txfonts}
\usepackage[pdfencoding=auto,psdextra]{hyperref}
\hypersetup{
    colorlinks=true,
    linkcolor=blue,
    filecolor=magenta,      
    urlcolor=blue,
    citecolor=blue
}
\makeatletter
\renewcommand*\aa@pageof{, page \thepage{} of \pageref*{LastPage}}
\makeatother

\usepackage[utf8]{inputenc}

\usepackage[switch, modulo]{lineno}

\usepackage{euclid}

\begin{document}
%
%
\title{\Euclid\/ preparation}
\subtitle{LXXXI. The impact of nonparametric star formation histories on spatially resolved galaxy property estimation using synthetic \Euclid images}


\newcommand{\orcid}[1]{} 
\author{Euclid Collaboration: A.~Nersesian\orcid{0000-0001-6843-409X}\thanks{\email{angelos.nersesian@ugent.be}}\inst{\ref{aff1},\ref{aff2}}
\and Abdurro'uf\orcid{0000-0002-5258-8761}\inst{\ref{aff3}}
\and M.~Baes\orcid{0000-0002-3930-2757}\inst{\ref{aff2}}
\and C.~Tortora\orcid{0000-0001-7958-6531}\inst{\ref{aff4}}
\and I.~Kova{\v{c}}i{\'{c}}\orcid{0000-0001-6751-3263}\inst{\ref{aff2}}
\and L.~Bisigello\orcid{0000-0003-0492-4924}\inst{\ref{aff5}}
\and P.~Corcho-Caballero\orcid{0000-0001-6327-7080}\inst{\ref{aff6}}
\and E.~Dur\'an-Camacho\orcid{0000-0002-3153-0536}\inst{\ref{aff7},\ref{aff8}}
\and L.~K.~Hunt\orcid{0000-0001-9162-2371}\inst{\ref{aff9}}
\and P.~Iglesias-Navarro\orcid{0009-0009-8959-2404}\inst{\ref{aff7},\ref{aff10}}
\and R.~Ragusa\inst{\ref{aff4}}
\and J.~Rom\'an\orcid{0000-0002-3849-3467}\inst{\ref{aff11}}
\and F.~Shankar\orcid{0000-0001-8973-5051}\inst{\ref{aff12}}
\and M.~Siudek\orcid{0000-0002-2949-2155}\inst{\ref{aff8},\ref{aff13}}
\and J.~G.~Sorce\orcid{0000-0002-2307-2432}\inst{\ref{aff14},\ref{aff15}}
\and F.~R.~Marleau\orcid{0000-0002-1442-2947}\inst{\ref{aff16}}
\and N.~Aghanim\orcid{0000-0002-6688-8992}\inst{\ref{aff15}}
\and S.~Andreon\orcid{0000-0002-2041-8784}\inst{\ref{aff17}}
\and N.~Auricchio\orcid{0000-0003-4444-8651}\inst{\ref{aff18}}
\and C.~Baccigalupi\orcid{0000-0002-8211-1630}\inst{\ref{aff19},\ref{aff20},\ref{aff21},\ref{aff22}}
\and M.~Baldi\orcid{0000-0003-4145-1943}\inst{\ref{aff23},\ref{aff18},\ref{aff24}}
\and S.~Bardelli\orcid{0000-0002-8900-0298}\inst{\ref{aff18}}
\and A.~Biviano\orcid{0000-0002-0857-0732}\inst{\ref{aff20},\ref{aff19}}
\and E.~Branchini\orcid{0000-0002-0808-6908}\inst{\ref{aff25},\ref{aff26},\ref{aff17}}
\and M.~Brescia\orcid{0000-0001-9506-5680}\inst{\ref{aff27},\ref{aff4}}
\and S.~Camera\orcid{0000-0003-3399-3574}\inst{\ref{aff28},\ref{aff29},\ref{aff30}}
\and G.~Ca\~nas-Herrera\orcid{0000-0003-2796-2149}\inst{\ref{aff31},\ref{aff32},\ref{aff33}}
\and V.~Capobianco\orcid{0000-0002-3309-7692}\inst{\ref{aff30}}
\and C.~Carbone\orcid{0000-0003-0125-3563}\inst{\ref{aff34}}
\and J.~Carretero\orcid{0000-0002-3130-0204}\inst{\ref{aff35},\ref{aff36}}
\and S.~Casas\orcid{0000-0002-4751-5138}\inst{\ref{aff37}}
\and M.~Castellano\orcid{0000-0001-9875-8263}\inst{\ref{aff38}}
\and G.~Castignani\orcid{0000-0001-6831-0687}\inst{\ref{aff18}}
\and S.~Cavuoti\orcid{0000-0002-3787-4196}\inst{\ref{aff4},\ref{aff39}}
\and A.~Cimatti\inst{\ref{aff40}}
\and C.~Colodro-Conde\inst{\ref{aff7}}
\and G.~Congedo\orcid{0000-0003-2508-0046}\inst{\ref{aff41}}
\and C.~J.~Conselice\orcid{0000-0003-1949-7638}\inst{\ref{aff42}}
\and L.~Conversi\orcid{0000-0002-6710-8476}\inst{\ref{aff43},\ref{aff44}}
\and Y.~Copin\orcid{0000-0002-5317-7518}\inst{\ref{aff45}}
\and F.~Courbin\orcid{0000-0003-0758-6510}\inst{\ref{aff46},\ref{aff47}}
\and H.~M.~Courtois\orcid{0000-0003-0509-1776}\inst{\ref{aff48}}
\and A.~Da~Silva\orcid{0000-0002-6385-1609}\inst{\ref{aff49},\ref{aff50}}
\and H.~Degaudenzi\orcid{0000-0002-5887-6799}\inst{\ref{aff51}}
\and G.~De~Lucia\orcid{0000-0002-6220-9104}\inst{\ref{aff20}}
\and H.~Dole\orcid{0000-0002-9767-3839}\inst{\ref{aff15}}
\and M.~Douspis\orcid{0000-0003-4203-3954}\inst{\ref{aff15}}
\and F.~Dubath\orcid{0000-0002-6533-2810}\inst{\ref{aff51}}
\and X.~Dupac\inst{\ref{aff44}}
\and S.~Dusini\orcid{0000-0002-1128-0664}\inst{\ref{aff52}}
\and M.~Farina\orcid{0000-0002-3089-7846}\inst{\ref{aff53}}
\and R.~Farinelli\inst{\ref{aff18}}
\and F.~Faustini\orcid{0000-0001-6274-5145}\inst{\ref{aff38},\ref{aff54}}
\and S.~Ferriol\inst{\ref{aff45}}
\and F.~Finelli\orcid{0000-0002-6694-3269}\inst{\ref{aff18},\ref{aff55}}
\and N.~Fourmanoit\orcid{0009-0005-6816-6925}\inst{\ref{aff56}}
\and M.~Frailis\orcid{0000-0002-7400-2135}\inst{\ref{aff20}}
\and E.~Franceschi\orcid{0000-0002-0585-6591}\inst{\ref{aff18}}
\and M.~Fumana\orcid{0000-0001-6787-5950}\inst{\ref{aff34}}
\and S.~Galeotta\orcid{0000-0002-3748-5115}\inst{\ref{aff20}}
\and K.~George\orcid{0000-0002-1734-8455}\inst{\ref{aff57}}
\and B.~Gillis\orcid{0000-0002-4478-1270}\inst{\ref{aff41}}
\and C.~Giocoli\orcid{0000-0002-9590-7961}\inst{\ref{aff18},\ref{aff24}}
\and J.~Gracia-Carpio\inst{\ref{aff58}}
\and A.~Grazian\orcid{0000-0002-5688-0663}\inst{\ref{aff5}}
\and F.~Grupp\inst{\ref{aff58},\ref{aff57}}
\and S.~V.~H.~Haugan\orcid{0000-0001-9648-7260}\inst{\ref{aff59}}
\and W.~Holmes\inst{\ref{aff60}}
\and F.~Hormuth\inst{\ref{aff61}}
\and A.~Hornstrup\orcid{0000-0002-3363-0936}\inst{\ref{aff62},\ref{aff63}}
\and K.~Jahnke\orcid{0000-0003-3804-2137}\inst{\ref{aff64}}
\and M.~Jhabvala\inst{\ref{aff65}}
\and E.~Keih\"anen\orcid{0000-0003-1804-7715}\inst{\ref{aff66}}
\and S.~Kermiche\orcid{0000-0002-0302-5735}\inst{\ref{aff56}}
\and M.~Kilbinger\orcid{0000-0001-9513-7138}\inst{\ref{aff67}}
\and B.~Kubik\orcid{0009-0006-5823-4880}\inst{\ref{aff45}}
\and M.~K\"ummel\orcid{0000-0003-2791-2117}\inst{\ref{aff57}}
\and M.~Kunz\orcid{0000-0002-3052-7394}\inst{\ref{aff68}}
\and H.~Kurki-Suonio\orcid{0000-0002-4618-3063}\inst{\ref{aff69},\ref{aff70}}
\and A.~M.~C.~Le~Brun\orcid{0000-0002-0936-4594}\inst{\ref{aff71}}
\and S.~Ligori\orcid{0000-0003-4172-4606}\inst{\ref{aff30}}
\and P.~B.~Lilje\orcid{0000-0003-4324-7794}\inst{\ref{aff59}}
\and V.~Lindholm\orcid{0000-0003-2317-5471}\inst{\ref{aff69},\ref{aff70}}
\and I.~Lloro\orcid{0000-0001-5966-1434}\inst{\ref{aff72}}
\and G.~Mainetti\orcid{0000-0003-2384-2377}\inst{\ref{aff73}}
\and D.~Maino\inst{\ref{aff74},\ref{aff34},\ref{aff75}}
\and E.~Maiorano\orcid{0000-0003-2593-4355}\inst{\ref{aff18}}
\and O.~Mansutti\orcid{0000-0001-5758-4658}\inst{\ref{aff20}}
\and O.~Marggraf\orcid{0000-0001-7242-3852}\inst{\ref{aff76}}
\and M.~Martinelli\orcid{0000-0002-6943-7732}\inst{\ref{aff38},\ref{aff77}}
\and N.~Martinet\orcid{0000-0003-2786-7790}\inst{\ref{aff78}}
\and F.~Marulli\orcid{0000-0002-8850-0303}\inst{\ref{aff79},\ref{aff18},\ref{aff24}}
\and R.~J.~Massey\orcid{0000-0002-6085-3780}\inst{\ref{aff80}}
\and E.~Medinaceli\orcid{0000-0002-4040-7783}\inst{\ref{aff18}}
\and S.~Mei\orcid{0000-0002-2849-559X}\inst{\ref{aff81},\ref{aff82}}
\and M.~Melchior\inst{\ref{aff83}}
\and Y.~Mellier\inst{\ref{aff84},\ref{aff85}}
\and M.~Meneghetti\orcid{0000-0003-1225-7084}\inst{\ref{aff18},\ref{aff24}}
\and E.~Merlin\orcid{0000-0001-6870-8900}\inst{\ref{aff38}}
\and G.~Meylan\inst{\ref{aff86}}
\and A.~Mora\orcid{0000-0002-1922-8529}\inst{\ref{aff87}}
\and M.~Moresco\orcid{0000-0002-7616-7136}\inst{\ref{aff79},\ref{aff18}}
\and L.~Moscardini\orcid{0000-0002-3473-6716}\inst{\ref{aff79},\ref{aff18},\ref{aff24}}
\and C.~Neissner\orcid{0000-0001-8524-4968}\inst{\ref{aff88},\ref{aff36}}
\and S.-M.~Niemi\orcid{0009-0005-0247-0086}\inst{\ref{aff31}}
\and C.~Padilla\orcid{0000-0001-7951-0166}\inst{\ref{aff88}}
\and S.~Paltani\orcid{0000-0002-8108-9179}\inst{\ref{aff51}}
\and F.~Pasian\orcid{0000-0002-4869-3227}\inst{\ref{aff20}}
\and K.~Pedersen\inst{\ref{aff89}}
\and V.~Pettorino\inst{\ref{aff31}}
\and S.~Pires\orcid{0000-0002-0249-2104}\inst{\ref{aff67}}
\and G.~Polenta\orcid{0000-0003-4067-9196}\inst{\ref{aff54}}
\and M.~Poncet\inst{\ref{aff90}}
\and L.~A.~Popa\inst{\ref{aff91}}
\and L.~Pozzetti\orcid{0000-0001-7085-0412}\inst{\ref{aff18}}
\and A.~Renzi\orcid{0000-0001-9856-1970}\inst{\ref{aff92},\ref{aff52}}
\and J.~Rhodes\orcid{0000-0002-4485-8549}\inst{\ref{aff60}}
\and G.~Riccio\inst{\ref{aff4}}
\and E.~Romelli\orcid{0000-0003-3069-9222}\inst{\ref{aff20}}
\and M.~Roncarelli\orcid{0000-0001-9587-7822}\inst{\ref{aff18}}
\and R.~Saglia\orcid{0000-0003-0378-7032}\inst{\ref{aff57},\ref{aff58}}
\and Z.~Sakr\orcid{0000-0002-4823-3757}\inst{\ref{aff93},\ref{aff94},\ref{aff95}}
\and D.~Sapone\orcid{0000-0001-7089-4503}\inst{\ref{aff96}}
\and B.~Sartoris\orcid{0000-0003-1337-5269}\inst{\ref{aff57},\ref{aff20}}
\and P.~Schneider\orcid{0000-0001-8561-2679}\inst{\ref{aff76}}
\and T.~Schrabback\orcid{0000-0002-6987-7834}\inst{\ref{aff16}}
\and A.~Secroun\orcid{0000-0003-0505-3710}\inst{\ref{aff56}}
\and G.~Seidel\orcid{0000-0003-2907-353X}\inst{\ref{aff64}}
\and M.~Seiffert\orcid{0000-0002-7536-9393}\inst{\ref{aff60}}
\and S.~Serrano\orcid{0000-0002-0211-2861}\inst{\ref{aff97},\ref{aff98},\ref{aff13}}
\and P.~Simon\inst{\ref{aff76}}
\and C.~Sirignano\orcid{0000-0002-0995-7146}\inst{\ref{aff92},\ref{aff52}}
\and G.~Sirri\orcid{0000-0003-2626-2853}\inst{\ref{aff24}}
\and L.~Stanco\orcid{0000-0002-9706-5104}\inst{\ref{aff52}}
\and J.~Steinwagner\orcid{0000-0001-7443-1047}\inst{\ref{aff58}}
\and P.~Tallada-Cresp\'{i}\orcid{0000-0002-1336-8328}\inst{\ref{aff35},\ref{aff36}}
\and A.~N.~Taylor\inst{\ref{aff41}}
\and I.~Tereno\orcid{0000-0002-4537-6218}\inst{\ref{aff49},\ref{aff99}}
\and S.~Toft\orcid{0000-0003-3631-7176}\inst{\ref{aff100},\ref{aff101}}
\and R.~Toledo-Moreo\orcid{0000-0002-2997-4859}\inst{\ref{aff102}}
\and F.~Torradeflot\orcid{0000-0003-1160-1517}\inst{\ref{aff36},\ref{aff35}}
\and I.~Tutusaus\orcid{0000-0002-3199-0399}\inst{\ref{aff94}}
\and L.~Valenziano\orcid{0000-0002-1170-0104}\inst{\ref{aff18},\ref{aff55}}
\and J.~Valiviita\orcid{0000-0001-6225-3693}\inst{\ref{aff69},\ref{aff70}}
\and T.~Vassallo\orcid{0000-0001-6512-6358}\inst{\ref{aff57},\ref{aff20}}
\and G.~Verdoes~Kleijn\orcid{0000-0001-5803-2580}\inst{\ref{aff6}}
\and A.~Veropalumbo\orcid{0000-0003-2387-1194}\inst{\ref{aff17},\ref{aff26},\ref{aff25}}
\and Y.~Wang\orcid{0000-0002-4749-2984}\inst{\ref{aff103}}
\and J.~Weller\orcid{0000-0002-8282-2010}\inst{\ref{aff57},\ref{aff58}}
\and G.~Zamorani\orcid{0000-0002-2318-301X}\inst{\ref{aff18}}
\and F.~M.~Zerbi\inst{\ref{aff17}}
\and I.~A.~Zinchenko\orcid{0000-0002-2944-2449}\inst{\ref{aff57}}
\and E.~Zucca\orcid{0000-0002-5845-8132}\inst{\ref{aff18}}
\and V.~Allevato\orcid{0000-0001-7232-5152}\inst{\ref{aff4}}
\and M.~Bolzonella\orcid{0000-0003-3278-4607}\inst{\ref{aff18}}
\and E.~Bozzo\orcid{0000-0002-8201-1525}\inst{\ref{aff51}}
\and C.~Burigana\orcid{0000-0002-3005-5796}\inst{\ref{aff104},\ref{aff55}}
\and R.~Cabanac\orcid{0000-0001-6679-2600}\inst{\ref{aff94}}
\and M.~Calabrese\orcid{0000-0002-2637-2422}\inst{\ref{aff105},\ref{aff34}}
\and A.~Cappi\inst{\ref{aff18},\ref{aff106}}
\and J.~A.~Escartin~Vigo\inst{\ref{aff58}}
\and L.~Gabarra\orcid{0000-0002-8486-8856}\inst{\ref{aff107}}
\and J.~Mart\'{i}n-Fleitas\orcid{0000-0002-8594-569X}\inst{\ref{aff108}}
\and S.~Matthew\orcid{0000-0001-8448-1697}\inst{\ref{aff41}}
\and N.~Mauri\orcid{0000-0001-8196-1548}\inst{\ref{aff40},\ref{aff24}}
\and R.~B.~Metcalf\orcid{0000-0003-3167-2574}\inst{\ref{aff79},\ref{aff18}}
\and A.~A.~Nucita\inst{\ref{aff109},\ref{aff110},\ref{aff111}}
\and A.~Pezzotta\orcid{0000-0003-0726-2268}\inst{\ref{aff17}}
\and M.~P\"ontinen\orcid{0000-0001-5442-2530}\inst{\ref{aff69}}
\and C.~Porciani\orcid{0000-0002-7797-2508}\inst{\ref{aff76}}
\and I.~Risso\orcid{0000-0003-2525-7761}\inst{\ref{aff17},\ref{aff26}}
\and V.~Scottez\orcid{0009-0008-3864-940X}\inst{\ref{aff84},\ref{aff112}}
\and M.~Sereno\orcid{0000-0003-0302-0325}\inst{\ref{aff18},\ref{aff24}}
\and M.~Tenti\orcid{0000-0002-4254-5901}\inst{\ref{aff24}}
\and M.~Viel\orcid{0000-0002-2642-5707}\inst{\ref{aff19},\ref{aff20},\ref{aff22},\ref{aff21},\ref{aff113}}
\and M.~Wiesmann\orcid{0009-0000-8199-5860}\inst{\ref{aff59}}
\and Y.~Akrami\orcid{0000-0002-2407-7956}\inst{\ref{aff114},\ref{aff115}}
\and I.~T.~Andika\orcid{0000-0001-6102-9526}\inst{\ref{aff116},\ref{aff117}}
\and S.~Anselmi\orcid{0000-0002-3579-9583}\inst{\ref{aff52},\ref{aff92},\ref{aff118}}
\and M.~Archidiacono\orcid{0000-0003-4952-9012}\inst{\ref{aff74},\ref{aff75}}
\and F.~Atrio-Barandela\orcid{0000-0002-2130-2513}\inst{\ref{aff119}}
\and D.~Bertacca\orcid{0000-0002-2490-7139}\inst{\ref{aff92},\ref{aff5},\ref{aff52}}
\and M.~Bethermin\orcid{0000-0002-3915-2015}\inst{\ref{aff120}}
\and A.~Blanchard\orcid{0000-0001-8555-9003}\inst{\ref{aff94}}
\and L.~Blot\orcid{0000-0002-9622-7167}\inst{\ref{aff121},\ref{aff71}}
\and S.~Borgani\orcid{0000-0001-6151-6439}\inst{\ref{aff122},\ref{aff19},\ref{aff20},\ref{aff21},\ref{aff113}}
\and M.~L.~Brown\orcid{0000-0002-0370-8077}\inst{\ref{aff42}}
\and S.~Bruton\orcid{0000-0002-6503-5218}\inst{\ref{aff123}}
\and A.~Calabro\orcid{0000-0003-2536-1614}\inst{\ref{aff38}}
\and B.~Camacho~Quevedo\orcid{0000-0002-8789-4232}\inst{\ref{aff19},\ref{aff22},\ref{aff20},\ref{aff13},\ref{aff97}}
\and F.~Caro\inst{\ref{aff38}}
\and C.~S.~Carvalho\inst{\ref{aff99}}
\and T.~Castro\orcid{0000-0002-6292-3228}\inst{\ref{aff20},\ref{aff21},\ref{aff19},\ref{aff113}}
\and F.~Cogato\orcid{0000-0003-4632-6113}\inst{\ref{aff79},\ref{aff18}}
\and S.~Conseil\orcid{0000-0002-3657-4191}\inst{\ref{aff45}}
\and A.~R.~Cooray\orcid{0000-0002-3892-0190}\inst{\ref{aff124}}
\and O.~Cucciati\orcid{0000-0002-9336-7551}\inst{\ref{aff18}}
\and S.~Davini\orcid{0000-0003-3269-1718}\inst{\ref{aff26}}
\and F.~De~Paolis\orcid{0000-0001-6460-7563}\inst{\ref{aff109},\ref{aff110},\ref{aff111}}
\and G.~Desprez\orcid{0000-0001-8325-1742}\inst{\ref{aff6}}
\and A.~D\'iaz-S\'anchez\orcid{0000-0003-0748-4768}\inst{\ref{aff125}}
\and J.~J.~Diaz\orcid{0000-0003-2101-1078}\inst{\ref{aff7}}
\and S.~Di~Domizio\orcid{0000-0003-2863-5895}\inst{\ref{aff25},\ref{aff26}}
\and J.~M.~Diego\orcid{0000-0001-9065-3926}\inst{\ref{aff126}}
\and P.-A.~Duc\orcid{0000-0003-3343-6284}\inst{\ref{aff120}}
\and A.~Enia\orcid{0000-0002-0200-2857}\inst{\ref{aff23},\ref{aff18}}
\and Y.~Fang\inst{\ref{aff57}}
\and A.~G.~Ferrari\orcid{0009-0005-5266-4110}\inst{\ref{aff24}}
\and A.~Finoguenov\orcid{0000-0002-4606-5403}\inst{\ref{aff69}}
\and A.~Fontana\orcid{0000-0003-3820-2823}\inst{\ref{aff38}}
\and F.~Fontanot\orcid{0000-0003-4744-0188}\inst{\ref{aff20},\ref{aff19}}
\and A.~Franco\orcid{0000-0002-4761-366X}\inst{\ref{aff110},\ref{aff109},\ref{aff111}}
\and K.~Ganga\orcid{0000-0001-8159-8208}\inst{\ref{aff81}}
\and J.~Garc\'ia-Bellido\orcid{0000-0002-9370-8360}\inst{\ref{aff114}}
\and T.~Gasparetto\orcid{0000-0002-7913-4866}\inst{\ref{aff20}}
\and V.~Gautard\inst{\ref{aff127}}
\and E.~Gaztanaga\orcid{0000-0001-9632-0815}\inst{\ref{aff13},\ref{aff97},\ref{aff128}}
\and F.~Giacomini\orcid{0000-0002-3129-2814}\inst{\ref{aff24}}
\and F.~Gianotti\orcid{0000-0003-4666-119X}\inst{\ref{aff18}}
\and G.~Gozaliasl\orcid{0000-0002-0236-919X}\inst{\ref{aff129},\ref{aff69}}
\and C.~M.~Gutierrez\orcid{0000-0001-7854-783X}\inst{\ref{aff8}}
\and A.~Hall\orcid{0000-0002-3139-8651}\inst{\ref{aff41}}
\and S.~Hemmati\orcid{0000-0003-2226-5395}\inst{\ref{aff130}}
\and C.~Hern\'andez-Monteagudo\orcid{0000-0001-5471-9166}\inst{\ref{aff10},\ref{aff7}}
\and H.~Hildebrandt\orcid{0000-0002-9814-3338}\inst{\ref{aff131}}
\and J.~Hjorth\orcid{0000-0002-4571-2306}\inst{\ref{aff89}}
\and J.~J.~E.~Kajava\orcid{0000-0002-3010-8333}\inst{\ref{aff132},\ref{aff133}}
\and Y.~Kang\orcid{0009-0000-8588-7250}\inst{\ref{aff51}}
\and V.~Kansal\orcid{0000-0002-4008-6078}\inst{\ref{aff134},\ref{aff135}}
\and D.~Karagiannis\orcid{0000-0002-4927-0816}\inst{\ref{aff136},\ref{aff137}}
\and K.~Kiiveri\inst{\ref{aff66}}
\and C.~C.~Kirkpatrick\inst{\ref{aff66}}
\and S.~Kruk\orcid{0000-0001-8010-8879}\inst{\ref{aff44}}
\and L.~Legrand\orcid{0000-0003-0610-5252}\inst{\ref{aff138},\ref{aff139}}
\and F.~Lepori\orcid{0009-0000-5061-7138}\inst{\ref{aff140}}
\and G.~Leroy\orcid{0009-0004-2523-4425}\inst{\ref{aff141},\ref{aff80}}
\and G.~F.~Lesci\orcid{0000-0002-4607-2830}\inst{\ref{aff79},\ref{aff18}}
\and J.~Lesgourgues\orcid{0000-0001-7627-353X}\inst{\ref{aff37}}
\and L.~Leuzzi\orcid{0009-0006-4479-7017}\inst{\ref{aff18}}
\and T.~I.~Liaudat\orcid{0000-0002-9104-314X}\inst{\ref{aff142}}
\and A.~Loureiro\orcid{0000-0002-4371-0876}\inst{\ref{aff143},\ref{aff144}}
\and J.~Macias-Perez\orcid{0000-0002-5385-2763}\inst{\ref{aff145}}
\and G.~Maggio\orcid{0000-0003-4020-4836}\inst{\ref{aff20}}
\and M.~Magliocchetti\orcid{0000-0001-9158-4838}\inst{\ref{aff53}}
\and F.~Mannucci\orcid{0000-0002-4803-2381}\inst{\ref{aff9}}
\and R.~Maoli\orcid{0000-0002-6065-3025}\inst{\ref{aff146},\ref{aff38}}
\and C.~J.~A.~P.~Martins\orcid{0000-0002-4886-9261}\inst{\ref{aff147},\ref{aff148}}
\and L.~Maurin\orcid{0000-0002-8406-0857}\inst{\ref{aff15}}
\and P.~Monaco\orcid{0000-0003-2083-7564}\inst{\ref{aff122},\ref{aff20},\ref{aff21},\ref{aff19}}
\and C.~Moretti\orcid{0000-0003-3314-8936}\inst{\ref{aff22},\ref{aff113},\ref{aff20},\ref{aff19},\ref{aff21}}
\and G.~Morgante\inst{\ref{aff18}}
\and K.~Naidoo\orcid{0000-0002-9182-1802}\inst{\ref{aff128}}
\and A.~Navarro-Alsina\orcid{0000-0002-3173-2592}\inst{\ref{aff76}}
\and F.~Passalacqua\orcid{0000-0002-8606-4093}\inst{\ref{aff92},\ref{aff52}}
\and K.~Paterson\orcid{0000-0001-8340-3486}\inst{\ref{aff64}}
\and L.~Patrizii\inst{\ref{aff24}}
\and A.~Pisani\orcid{0000-0002-6146-4437}\inst{\ref{aff56}}
\and D.~Potter\orcid{0000-0002-0757-5195}\inst{\ref{aff140}}
\and S.~Quai\orcid{0000-0002-0449-8163}\inst{\ref{aff79},\ref{aff18}}
\and M.~Radovich\orcid{0000-0002-3585-866X}\inst{\ref{aff5}}
\and P.-F.~Rocci\inst{\ref{aff15}}
\and G.~Rodighiero\orcid{0000-0002-9415-2296}\inst{\ref{aff92},\ref{aff5}}
\and S.~Sacquegna\orcid{0000-0002-8433-6630}\inst{\ref{aff109},\ref{aff110},\ref{aff149}}
\and M.~Sahl\'en\orcid{0000-0003-0973-4804}\inst{\ref{aff150}}
\and D.~B.~Sanders\orcid{0000-0002-1233-9998}\inst{\ref{aff151}}
\and E.~Sarpa\orcid{0000-0002-1256-655X}\inst{\ref{aff22},\ref{aff113},\ref{aff21}}
\and A.~Schneider\orcid{0000-0001-7055-8104}\inst{\ref{aff140}}
\and M.~Schultheis\inst{\ref{aff106}}
\and D.~Sciotti\orcid{0009-0008-4519-2620}\inst{\ref{aff38},\ref{aff77}}
\and E.~Sellentin\inst{\ref{aff152},\ref{aff33}}
\and L.~C.~Smith\orcid{0000-0002-3259-2771}\inst{\ref{aff153}}
\and K.~Tanidis\orcid{0000-0001-9843-5130}\inst{\ref{aff107}}
\and C.~Tao\orcid{0000-0001-7961-8177}\inst{\ref{aff56}}
\and G.~Testera\inst{\ref{aff26}}
\and R.~Teyssier\orcid{0000-0001-7689-0933}\inst{\ref{aff154}}
\and S.~Tosi\orcid{0000-0002-7275-9193}\inst{\ref{aff25},\ref{aff26},\ref{aff17}}
\and A.~Troja\orcid{0000-0003-0239-4595}\inst{\ref{aff92},\ref{aff52}}
\and M.~Tucci\inst{\ref{aff51}}
\and C.~Valieri\inst{\ref{aff24}}
\and A.~Venhola\orcid{0000-0001-6071-4564}\inst{\ref{aff155}}
\and D.~Vergani\orcid{0000-0003-0898-2216}\inst{\ref{aff18}}
\and G.~Verza\orcid{0000-0002-1886-8348}\inst{\ref{aff156}}
\and P.~Vielzeuf\orcid{0000-0003-2035-9339}\inst{\ref{aff56}}
\and N.~A.~Walton\orcid{0000-0003-3983-8778}\inst{\ref{aff153}}
\and J.~H.~Knapen\orcid{0000-0003-1643-0024}\inst{\ref{aff7},\ref{aff10}}}
										   
\institute{STAR Institute, University of Li{\`e}ge, Quartier Agora, All\'ee du six Ao\^ut 19c, 4000 Li\`ege, Belgium\label{aff1}
\and
Sterrenkundig Observatorium, Universiteit Gent, Krijgslaan 281 S9, 9000 Gent, Belgium\label{aff2}
\and
Johns Hopkins University 3400 North Charles Street Baltimore, MD 21218, USA\label{aff3}
\and
INAF-Osservatorio Astronomico di Capodimonte, Via Moiariello 16, 80131 Napoli, Italy\label{aff4}
\and
INAF-Osservatorio Astronomico di Padova, Via dell'Osservatorio 5, 35122 Padova, Italy\label{aff5}
\and
Kapteyn Astronomical Institute, University of Groningen, PO Box 800, 9700 AV Groningen, The Netherlands\label{aff6}
\and
Instituto de Astrof\'{\i}sica de Canarias, E-38205 La Laguna, Tenerife, Spain\label{aff7}
\and
 Instituto de Astrof\'{\i}sica de Canarias, E-38205 La Laguna; Universidad de La Laguna, Dpto. Astrof\'\i sica, E-38206 La Laguna, Tenerife, Spain\label{aff8}
\and
INAF-Osservatorio Astrofisico di Arcetri, Largo E. Fermi 5, 50125, Firenze, Italy\label{aff9}
\and
Universidad de La Laguna, Dpto. Astrof\'\i sica, E-38206 La Laguna, Tenerife, Spain\label{aff10}
\and
Departamento de F{\'\i}sica de la Tierra y Astrof{\'\i}sica, Universidad Complutense de Madrid, Plaza de las Ciencias 2, E-28040 Madrid, Spain\label{aff11}
\and
School of Physics \& Astronomy, University of Southampton, Highfield Campus, Southampton SO17 1BJ, UK\label{aff12}
\and
Institute of Space Sciences (ICE, CSIC), Campus UAB, Carrer de Can Magrans, s/n, 08193 Barcelona, Spain\label{aff13}
\and
Univ. Lille, CNRS, Centrale Lille, UMR 9189 CRIStAL, 59000 Lille, France\label{aff14}
\and
Universit\'e Paris-Saclay, CNRS, Institut d'astrophysique spatiale, 91405, Orsay, France\label{aff15}
\and
Universit\"at Innsbruck, Institut f\"ur Astro- und Teilchenphysik, Technikerstr. 25/8, 6020 Innsbruck, Austria\label{aff16}
\and
INAF-Osservatorio Astronomico di Brera, Via Brera 28, 20122 Milano, Italy\label{aff17}
\and
INAF-Osservatorio di Astrofisica e Scienza dello Spazio di Bologna, Via Piero Gobetti 93/3, 40129 Bologna, Italy\label{aff18}
\and
IFPU, Institute for Fundamental Physics of the Universe, via Beirut 2, 34151 Trieste, Italy\label{aff19}
\and
INAF-Osservatorio Astronomico di Trieste, Via G. B. Tiepolo 11, 34143 Trieste, Italy\label{aff20}
\and
INFN, Sezione di Trieste, Via Valerio 2, 34127 Trieste TS, Italy\label{aff21}
\and
SISSA, International School for Advanced Studies, Via Bonomea 265, 34136 Trieste TS, Italy\label{aff22}
\and
Dipartimento di Fisica e Astronomia, Universit\`a di Bologna, Via Gobetti 93/2, 40129 Bologna, Italy\label{aff23}
\and
INFN-Sezione di Bologna, Viale Berti Pichat 6/2, 40127 Bologna, Italy\label{aff24}
\and
Dipartimento di Fisica, Universit\`a di Genova, Via Dodecaneso 33, 16146, Genova, Italy\label{aff25}
\and
INFN-Sezione di Genova, Via Dodecaneso 33, 16146, Genova, Italy\label{aff26}
\and
Department of Physics "E. Pancini", University Federico II, Via Cinthia 6, 80126, Napoli, Italy\label{aff27}
\and
Dipartimento di Fisica, Universit\`a degli Studi di Torino, Via P. Giuria 1, 10125 Torino, Italy\label{aff28}
\and
INFN-Sezione di Torino, Via P. Giuria 1, 10125 Torino, Italy\label{aff29}
\and
INAF-Osservatorio Astrofisico di Torino, Via Osservatorio 20, 10025 Pino Torinese (TO), Italy\label{aff30}
\and
European Space Agency/ESTEC, Keplerlaan 1, 2201 AZ Noordwijk, The Netherlands\label{aff31}
\and
Institute Lorentz, Leiden University, Niels Bohrweg 2, 2333 CA Leiden, The Netherlands\label{aff32}
\and
Leiden Observatory, Leiden University, Einsteinweg 55, 2333 CC Leiden, The Netherlands\label{aff33}
\and
INAF-IASF Milano, Via Alfonso Corti 12, 20133 Milano, Italy\label{aff34}
\and
Centro de Investigaciones Energ\'eticas, Medioambientales y Tecnol\'ogicas (CIEMAT), Avenida Complutense 40, 28040 Madrid, Spain\label{aff35}
\and
Port d'Informaci\'{o} Cient\'{i}fica, Campus UAB, C. Albareda s/n, 08193 Bellaterra (Barcelona), Spain\label{aff36}
\and
Institute for Theoretical Particle Physics and Cosmology (TTK), RWTH Aachen University, 52056 Aachen, Germany\label{aff37}
\and
INAF-Osservatorio Astronomico di Roma, Via Frascati 33, 00078 Monteporzio Catone, Italy\label{aff38}
\and
INFN section of Naples, Via Cinthia 6, 80126, Napoli, Italy\label{aff39}
\and
Dipartimento di Fisica e Astronomia "Augusto Righi" - Alma Mater Studiorum Universit\`a di Bologna, Viale Berti Pichat 6/2, 40127 Bologna, Italy\label{aff40}
\and
Institute for Astronomy, University of Edinburgh, Royal Observatory, Blackford Hill, Edinburgh EH9 3HJ, UK\label{aff41}
\and
Jodrell Bank Centre for Astrophysics, Department of Physics and Astronomy, University of Manchester, Oxford Road, Manchester M13 9PL, UK\label{aff42}
\and
European Space Agency/ESRIN, Largo Galileo Galilei 1, 00044 Frascati, Roma, Italy\label{aff43}
\and
ESAC/ESA, Camino Bajo del Castillo, s/n., Urb. Villafranca del Castillo, 28692 Villanueva de la Ca\~nada, Madrid, Spain\label{aff44}
\and
Universit\'e Claude Bernard Lyon 1, CNRS/IN2P3, IP2I Lyon, UMR 5822, Villeurbanne, F-69100, France\label{aff45}
\and
Institut de Ci\`{e}ncies del Cosmos (ICCUB), Universitat de Barcelona (IEEC-UB), Mart\'{i} i Franqu\`{e}s 1, 08028 Barcelona, Spain\label{aff46}
\and
Instituci\'o Catalana de Recerca i Estudis Avan\c{c}ats (ICREA), Passeig de Llu\'{\i}s Companys 23, 08010 Barcelona, Spain\label{aff47}
\and
UCB Lyon 1, CNRS/IN2P3, IUF, IP2I Lyon, 4 rue Enrico Fermi, 69622 Villeurbanne, France\label{aff48}
\and
Departamento de F\'isica, Faculdade de Ci\^encias, Universidade de Lisboa, Edif\'icio C8, Campo Grande, PT1749-016 Lisboa, Portugal\label{aff49}
\and
Instituto de Astrof\'isica e Ci\^encias do Espa\c{c}o, Faculdade de Ci\^encias, Universidade de Lisboa, Campo Grande, 1749-016 Lisboa, Portugal\label{aff50}
\and
Department of Astronomy, University of Geneva, ch. d'Ecogia 16, 1290 Versoix, Switzerland\label{aff51}
\and
INFN-Padova, Via Marzolo 8, 35131 Padova, Italy\label{aff52}
\and
INAF-Istituto di Astrofisica e Planetologia Spaziali, via del Fosso del Cavaliere, 100, 00100 Roma, Italy\label{aff53}
\and
Space Science Data Center, Italian Space Agency, via del Politecnico snc, 00133 Roma, Italy\label{aff54}
\and
INFN-Bologna, Via Irnerio 46, 40126 Bologna, Italy\label{aff55}
\and
Aix-Marseille Universit\'e, CNRS/IN2P3, CPPM, Marseille, France\label{aff56}
\and
Universit\"ats-Sternwarte M\"unchen, Fakult\"at f\"ur Physik, Ludwig-Maximilians-Universit\"at M\"unchen, Scheinerstr.~1, 81679 M\"unchen, Germany\label{aff57}
\and
Max Planck Institute for Extraterrestrial Physics, Giessenbachstr. 1, 85748 Garching, Germany\label{aff58}
\and
Institute of Theoretical Astrophysics, University of Oslo, P.O. Box 1029 Blindern, 0315 Oslo, Norway\label{aff59}
\and
Jet Propulsion Laboratory, California Institute of Technology, 4800 Oak Grove Drive, Pasadena, CA, 91109, USA\label{aff60}
\and
Felix Hormuth Engineering, Goethestr. 17, 69181 Leimen, Germany\label{aff61}
\and
Technical University of Denmark, Elektrovej 327, 2800 Kgs. Lyngby, Denmark\label{aff62}
\and
Cosmic Dawn Center (DAWN), Denmark\label{aff63}
\and
Max-Planck-Institut f\"ur Astronomie, K\"onigstuhl 17, 69117 Heidelberg, Germany\label{aff64}
\and
NASA Goddard Space Flight Center, Greenbelt, MD 20771, USA\label{aff65}
\and
Department of Physics and Helsinki Institute of Physics, Gustaf H\"allstr\"omin katu 2, University of Helsinki, 00014 Helsinki, Finland\label{aff66}
\and
Universit\'e Paris-Saclay, Universit\'e Paris Cit\'e, CEA, CNRS, AIM, 91191, Gif-sur-Yvette, France\label{aff67}
\and
Universit\'e de Gen\`eve, D\'epartement de Physique Th\'eorique and Centre for Astroparticle Physics, 24 quai Ernest-Ansermet, CH-1211 Gen\`eve 4, Switzerland\label{aff68}
\and
Department of Physics, P.O. Box 64, University of Helsinki, 00014 Helsinki, Finland\label{aff69}
\and
Helsinki Institute of Physics, Gustaf H{\"a}llstr{\"o}min katu 2, University of Helsinki, 00014 Helsinki, Finland\label{aff70}
\and
Laboratoire d'etude de l'Univers et des phenomenes eXtremes, Observatoire de Paris, Universit\'e PSL, Sorbonne Universit\'e, CNRS, 92190 Meudon, France\label{aff71}
\and
SKAO, Jodrell Bank, Lower Withington, Macclesfield SK11 9FT, UK\label{aff72}
\and
Centre de Calcul de l'IN2P3/CNRS, 21 avenue Pierre de Coubertin 69627 Villeurbanne Cedex, France\label{aff73}
\and
Dipartimento di Fisica "Aldo Pontremoli", Universit\`a degli Studi di Milano, Via Celoria 16, 20133 Milano, Italy\label{aff74}
\and
INFN-Sezione di Milano, Via Celoria 16, 20133 Milano, Italy\label{aff75}
\and
Universit\"at Bonn, Argelander-Institut f\"ur Astronomie, Auf dem H\"ugel 71, 53121 Bonn, Germany\label{aff76}
\and
INFN-Sezione di Roma, Piazzale Aldo Moro, 2 - c/o Dipartimento di Fisica, Edificio G. Marconi, 00185 Roma, Italy\label{aff77}
\and
Aix-Marseille Universit\'e, CNRS, CNES, LAM, Marseille, France\label{aff78}
\and
Dipartimento di Fisica e Astronomia "Augusto Righi" - Alma Mater Studiorum Universit\`a di Bologna, via Piero Gobetti 93/2, 40129 Bologna, Italy\label{aff79}
\and
Department of Physics, Institute for Computational Cosmology, Durham University, South Road, Durham, DH1 3LE, UK\label{aff80}
\and
Universit\'e Paris Cit\'e, CNRS, Astroparticule et Cosmologie, 75013 Paris, France\label{aff81}
\and
CNRS-UCB International Research Laboratory, Centre Pierre Bin\'etruy, IRL2007, CPB-IN2P3, Berkeley, USA\label{aff82}
\and
University of Applied Sciences and Arts of Northwestern Switzerland, School of Engineering, 5210 Windisch, Switzerland\label{aff83}
\and
Institut d'Astrophysique de Paris, 98bis Boulevard Arago, 75014, Paris, France\label{aff84}
\and
Institut d'Astrophysique de Paris, UMR 7095, CNRS, and Sorbonne Universit\'e, 98 bis boulevard Arago, 75014 Paris, France\label{aff85}
\and
Institute of Physics, Laboratory of Astrophysics, Ecole Polytechnique F\'ed\'erale de Lausanne (EPFL), Observatoire de Sauverny, 1290 Versoix, Switzerland\label{aff86}
\and
Telespazio UK S.L. for European Space Agency (ESA), Camino bajo del Castillo, s/n, Urbanizacion Villafranca del Castillo, Villanueva de la Ca\~nada, 28692 Madrid, Spain\label{aff87}
\and
Institut de F\'{i}sica d'Altes Energies (IFAE), The Barcelona Institute of Science and Technology, Campus UAB, 08193 Bellaterra (Barcelona), Spain\label{aff88}
\and
DARK, Niels Bohr Institute, University of Copenhagen, Jagtvej 155, 2200 Copenhagen, Denmark\label{aff89}
\and
Centre National d'Etudes Spatiales -- Centre spatial de Toulouse, 18 avenue Edouard Belin, 31401 Toulouse Cedex 9, France\label{aff90}
\and
Institute of Space Science, Str. Atomistilor, nr. 409 M\u{a}gurele, Ilfov, 077125, Romania\label{aff91}
\and
Dipartimento di Fisica e Astronomia "G. Galilei", Universit\`a di Padova, Via Marzolo 8, 35131 Padova, Italy\label{aff92}
\and
Institut f\"ur Theoretische Physik, University of Heidelberg, Philosophenweg 16, 69120 Heidelberg, Germany\label{aff93}
\and
Institut de Recherche en Astrophysique et Plan\'etologie (IRAP), Universit\'e de Toulouse, CNRS, UPS, CNES, 14 Av. Edouard Belin, 31400 Toulouse, France\label{aff94}
\and
Universit\'e St Joseph; Faculty of Sciences, Beirut, Lebanon\label{aff95}
\and
Departamento de F\'isica, FCFM, Universidad de Chile, Blanco Encalada 2008, Santiago, Chile\label{aff96}
\and
Institut d'Estudis Espacials de Catalunya (IEEC),  Edifici RDIT, Campus UPC, 08860 Castelldefels, Barcelona, Spain\label{aff97}
\and
Satlantis, University Science Park, Sede Bld 48940, Leioa-Bilbao, Spain\label{aff98}
\and
Instituto de Astrof\'isica e Ci\^encias do Espa\c{c}o, Faculdade de Ci\^encias, Universidade de Lisboa, Tapada da Ajuda, 1349-018 Lisboa, Portugal\label{aff99}
\and
Cosmic Dawn Center (DAWN)\label{aff100}
\and
Niels Bohr Institute, University of Copenhagen, Jagtvej 128, 2200 Copenhagen, Denmark\label{aff101}
\and
Universidad Polit\'ecnica de Cartagena, Departamento de Electr\'onica y Tecnolog\'ia de Computadoras,  Plaza del Hospital 1, 30202 Cartagena, Spain\label{aff102}
\and
Infrared Processing and Analysis Center, California Institute of Technology, Pasadena, CA 91125, USA\label{aff103}
\and
INAF, Istituto di Radioastronomia, Via Piero Gobetti 101, 40129 Bologna, Italy\label{aff104}
\and
Astronomical Observatory of the Autonomous Region of the Aosta Valley (OAVdA), Loc. Lignan 39, I-11020, Nus (Aosta Valley), Italy\label{aff105}
\and
Universit\'e C\^{o}te d'Azur, Observatoire de la C\^{o}te d'Azur, CNRS, Laboratoire Lagrange, Bd de l'Observatoire, CS 34229, 06304 Nice cedex 4, France\label{aff106}
\and
Department of Physics, Oxford University, Keble Road, Oxford OX1 3RH, UK\label{aff107}
\and
Aurora Technology for European Space Agency (ESA), Camino bajo del Castillo, s/n, Urbanizacion Villafranca del Castillo, Villanueva de la Ca\~nada, 28692 Madrid, Spain\label{aff108}
\and
Department of Mathematics and Physics E. De Giorgi, University of Salento, Via per Arnesano, CP-I93, 73100, Lecce, Italy\label{aff109}
\and
INFN, Sezione di Lecce, Via per Arnesano, CP-193, 73100, Lecce, Italy\label{aff110}
\and
INAF-Sezione di Lecce, c/o Dipartimento Matematica e Fisica, Via per Arnesano, 73100, Lecce, Italy\label{aff111}
\and
ICL, Junia, Universit\'e Catholique de Lille, LITL, 59000 Lille, France\label{aff112}
\and
ICSC - Centro Nazionale di Ricerca in High Performance Computing, Big Data e Quantum Computing, Via Magnanelli 2, Bologna, Italy\label{aff113}
\and
Instituto de F\'isica Te\'orica UAM-CSIC, Campus de Cantoblanco, 28049 Madrid, Spain\label{aff114}
\and
CERCA/ISO, Department of Physics, Case Western Reserve University, 10900 Euclid Avenue, Cleveland, OH 44106, USA\label{aff115}
\and
Technical University of Munich, TUM School of Natural Sciences, Physics Department, James-Franck-Str.~1, 85748 Garching, Germany\label{aff116}
\and
Max-Planck-Institut f\"ur Astrophysik, Karl-Schwarzschild-Str.~1, 85748 Garching, Germany\label{aff117}
\and
Laboratoire Univers et Th\'eorie, Observatoire de Paris, Universit\'e PSL, Universit\'e Paris Cit\'e, CNRS, 92190 Meudon, France\label{aff118}
\and
Departamento de F{\'\i}sica Fundamental. Universidad de Salamanca. Plaza de la Merced s/n. 37008 Salamanca, Spain\label{aff119}
\and
Universit\'e de Strasbourg, CNRS, Observatoire astronomique de Strasbourg, UMR 7550, 67000 Strasbourg, France\label{aff120}
\and
Center for Data-Driven Discovery, Kavli IPMU (WPI), UTIAS, The University of Tokyo, Kashiwa, Chiba 277-8583, Japan\label{aff121}
\and
Dipartimento di Fisica - Sezione di Astronomia, Universit\`a di Trieste, Via Tiepolo 11, 34131 Trieste, Italy\label{aff122}
\and
California Institute of Technology, 1200 E California Blvd, Pasadena, CA 91125, USA\label{aff123}
\and
Department of Physics \& Astronomy, University of California Irvine, Irvine CA 92697, USA\label{aff124}
\and
Departamento F\'isica Aplicada, Universidad Polit\'ecnica de Cartagena, Campus Muralla del Mar, 30202 Cartagena, Murcia, Spain\label{aff125}
\and
Instituto de F\'isica de Cantabria, Edificio Juan Jord\'a, Avenida de los Castros, 39005 Santander, Spain\label{aff126}
\and
CEA Saclay, DFR/IRFU, Service d'Astrophysique, Bat. 709, 91191 Gif-sur-Yvette, France\label{aff127}
\and
Institute of Cosmology and Gravitation, University of Portsmouth, Portsmouth PO1 3FX, UK\label{aff128}
\and
Department of Computer Science, Aalto University, PO Box 15400, Espoo, FI-00 076, Finland\label{aff129}
\and
Caltech/IPAC, 1200 E. California Blvd., Pasadena, CA 91125, USA\label{aff130}
\and
Ruhr University Bochum, Faculty of Physics and Astronomy, Astronomical Institute (AIRUB), German Centre for Cosmological Lensing (GCCL), 44780 Bochum, Germany\label{aff131}
\and
Department of Physics and Astronomy, Vesilinnantie 5, University of Turku, 20014 Turku, Finland\label{aff132}
\and
Serco for European Space Agency (ESA), Camino bajo del Castillo, s/n, Urbanizacion Villafranca del Castillo, Villanueva de la Ca\~nada, 28692 Madrid, Spain\label{aff133}
\and
ARC Centre of Excellence for Dark Matter Particle Physics, Melbourne, Australia\label{aff134}
\and
Centre for Astrophysics \& Supercomputing, Swinburne University of Technology,  Hawthorn, Victoria 3122, Australia\label{aff135}
\and
Dipartimento di Fisica e Scienze della Terra, Universit\`a degli Studi di Ferrara, Via Giuseppe Saragat 1, 44122 Ferrara, Italy\label{aff136}
\and
Department of Physics and Astronomy, University of the Western Cape, Bellville, Cape Town, 7535, South Africa\label{aff137}
\and
DAMTP, Centre for Mathematical Sciences, Wilberforce Road, Cambridge CB3 0WA, UK\label{aff138}
\and
Kavli Institute for Cosmology Cambridge, Madingley Road, Cambridge, CB3 0HA, UK\label{aff139}
\and
Department of Astrophysics, University of Zurich, Winterthurerstrasse 190, 8057 Zurich, Switzerland\label{aff140}
\and
Department of Physics, Centre for Extragalactic Astronomy, Durham University, South Road, Durham, DH1 3LE, UK\label{aff141}
\and
IRFU, CEA, Universit\'e Paris-Saclay 91191 Gif-sur-Yvette Cedex, France\label{aff142}
\and
Oskar Klein Centre for Cosmoparticle Physics, Department of Physics, Stockholm University, Stockholm, SE-106 91, Sweden\label{aff143}
\and
Astrophysics Group, Blackett Laboratory, Imperial College London, London SW7 2AZ, UK\label{aff144}
\and
Univ. Grenoble Alpes, CNRS, Grenoble INP, LPSC-IN2P3, 53, Avenue des Martyrs, 38000, Grenoble, France\label{aff145}
\and
Dipartimento di Fisica, Sapienza Universit\`a di Roma, Piazzale Aldo Moro 2, 00185 Roma, Italy\label{aff146}
\and
Centro de Astrof\'{\i}sica da Universidade do Porto, Rua das Estrelas, 4150-762 Porto, Portugal\label{aff147}
\and
Instituto de Astrof\'isica e Ci\^encias do Espa\c{c}o, Universidade do Porto, CAUP, Rua das Estrelas, PT4150-762 Porto, Portugal\label{aff148}
\and
INAF - Osservatorio Astronomico d'Abruzzo, Via Maggini, 64100, Teramo, Italy\label{aff149}
\and
Theoretical astrophysics, Department of Physics and Astronomy, Uppsala University, Box 516, 751 37 Uppsala, Sweden\label{aff150}
\and
Institute for Astronomy, University of Hawaii, 2680 Woodlawn Drive, Honolulu, HI 96822, USA\label{aff151}
\and
Mathematical Institute, University of Leiden, Einsteinweg 55, 2333 CA Leiden, The Netherlands\label{aff152}
\and
Institute of Astronomy, University of Cambridge, Madingley Road, Cambridge CB3 0HA, UK\label{aff153}
\and
Department of Astrophysical Sciences, Peyton Hall, Princeton University, Princeton, NJ 08544, USA\label{aff154}
\and
Space physics and astronomy research unit, University of Oulu, Pentti Kaiteran katu 1, FI-90014 Oulu, Finland\label{aff155}
\and
Center for Computational Astrophysics, Flatiron Institute, 162 5th Avenue, 10010, New York, NY, USA\label{aff156}} 
%
\abstract
   {}
   {We analyzed the spatially resolved and global star formation histories (SFHs) for a sample of 25 TNG50-{\tt SKIRT} Atlas galaxies to assess the feasibility of reconstructing accurate SFHs from \Euclid-like data. This study provides a proof of concept for extracting the spatially resolved SFHs of local galaxies with \Euclid, highlighting the strengths and limitations of SFH modeling in the context of next-generation galaxy surveys.}
   {We used the spectral energy distribution (SED) fitting code {\tt Prospector} to model both spatially resolved and global SFHs using parametric and nonparametric configurations. The input consisted of mock ultraviolet--near-infrared photometry derived from the TNG50 cosmological simulation and processed with the radiative transfer code {\tt SKIRT}.}
   {We show that nonparametric SFHs provide a more effective approach to mitigating the outshining effect by recent star formation, offering improved accuracy in the determination of galaxy stellar properties. Also, we find that the nonparametric SFH model at resolved scales closely recovers the stellar mass formation times (within 0.1~dex) and the ground truth values from TNG50, with an absolute average bias of $0.03$~dex in stellar mass and $0.01$~dex in both specific star formation rate and mass-weighted age. In contrast, larger offsets are estimated for all stellar properties and formation times when using a simple $\tau$-model SFH, at both resolved and global scales, highlighting its limitations.}
   {These results emphasize the critical role of nonparametric SFHs in both global and spatially resolved analyses, as they better capture the complex evolutionary pathways of galaxies and avoid the biases inherent in simple parametric models.}
%
%
%
%
   \titlerunning{\Euclid preparation: Spatially resolved SFHs from synthetic \Euclid images}
   \authorrunning{Euclid Collaboration: A. Nersesian et al.}
   
   \maketitle
%
%
%
%
   
\section{\label{sc:Intro} Introduction}

Reconstructing the star formation histories (SFHs) of galaxies is a critical step toward deciphering galaxy evolution across cosmic time. An accurate reconstruction of a galaxy's SFH enables reliable estimates of its total stellar mass and star formation rate (SFR). Of course, there are multiple factors that regulate the process of star formation and influence the overall shape and amplitude of the SFH. First, environmental factors such as galaxy mergers or intense disk instabilities can induce episodes of star formation and active galactic nuclei (AGN) activity in the core \citep{Barnes_1991ApJ...370L..65B, Barnes_1996ApJ...471..115B, Springel_2005ApJ...620L..79S, Snyder_2011ApJ...741...77S, Dekel_2014MNRAS.438.1870D, Zolotov_2015MNRAS.450.2327Z, Wellons_2015MNRAS.449..361W, Lapiner_2023MNRAS.522.4515L}. Second, feedback mechanisms, such as supernova explosions and AGNs, can prevent the hot gas in a galaxy’s halo from cooling down \citep[e.g.,][]{Crenshaw_2003ARA&A..41..117C, Best_2005MNRAS.362...25B, Croton_2006MNRAS.365...11C, Fabian_2012ARA&A..50..455F, Cheung_2016Natur.533..504C, Harrison_2017NatAs...1E.165H, Barisic_2017ApJ...847...72B, Terrazas_2017ApJ...844..170T, Henden_2018MNRAS.479.5385H, Semenov_2021ApJ...918...13S, Dome_2024MNRAS.527.2139D}. Finally, dynamical features such as spiral arms and stellar bars can regulate the process of star formation locally \citep{Elmegreen_2011EAS....51...19E, Baba_2013ApJ...763...46B, Dobbs_2014PASA...31...35D, Selwood_2019MNRAS.489..116S, Pettitt_2020MNRAS.498.1159P, Shin_2023ApJ...947...61S, Duran_Camacho_2025arXiv250613560D}. Therefore, the study of SFHs not only sheds light on when and how galaxies formed their stars but also helps link their past to present-day properties, such as their morphology, stellar populations, and chemical composition.

The most widely used method for inferring the SFHs of individual galaxies is modeling their observed spectral energy distribution (SED). This approach usually involves a set of multiwavelength photometric measurements that is compared to a library of theoretical and empirical models of galaxy evolution \citep[see reviews by][]{Walcher_2011Ap&SS.331....1W, Conroy_2013ARA&A..51..393C, Pacifici_2023ApJ...944..141P}. Ideally, broadband photometry should be combined with spectroscopic information to reduce the uncertainties in the SFHs of galaxies, particularly at earlier look-back times \citep[e.g.,][]{Worthey_1994ApJS...94..687W, Bruzual_2003MNRAS.344.1000B, Trager_2000AJ....120..165T, Gallazzi_2009ApJS..185..253G, Salim_2016ApJS..227....2S, Salim_2018ApJ...859...11S, Chauke_2018ApJ...861...13C, Johnson_2021ApJS..254...22J, Thorne_2021MNRAS.505..540T, Tacchella_2022ApJ...926..134T, Nersesian_2024A&A...681A..94N, Iglesias_Navarro_2024A&A...689A..58I, Csizi_2024A&A...689A..37C, Siudek_2024A&A...691A.308S, Kaushal_2024ApJ...961..118K, Nersesian_2025A&A...695A..86N}. Nevertheless, in the local Universe, reconstructing the stellar age distribution of individual galaxies through high-spectral-resolution spectroscopy remains a significant challenge, as most galaxies are older than 5~Gyr \citep{Gallazzi_2005MNRAS.362...41G}, resulting in stellar spectra that are often very similar in terms of absorption features.

One of the assumptions that goes into SED fitting is the choice of the prior function for the SFH. This choice can significantly influence the reliability and precision of the derived physical properties of galaxies \citep[e.g.,][]{Carnall_2019ApJ...873...44C, Leja_2019ApJ...876....3L, Siudek_2024A&A...691A.308S}. Most SED fitting frameworks adopt simplified parametric SFH models to describe the evolution of SFR with time. These models typically include parameters that control normalization, amplitudes, and characteristic timescales (e.g., the $e$-folding time). Common examples include exponentially declining or rising SFHs, lognormal forms, and delayed-$\tau$ models \citep{Papovich_2011MNRAS.412.1123P, Diemer_2017ApJ...839...26D, Boquien_2019A&A...622A.103B}. More flexible parametric models also exist, allowing for multiple episodes of star formation and quenching \citep{Ciesla_2017A&A...608A..41C, Morishita_2019ApJ...877..141M, Abdurrouf_2021ApJS..254...15A}.

While these models are computationally less intensive, their fixed functional forms often fail to capture the expected diversity of SFHs. Numerous studies have demonstrated that this limitation can introduce systematic biases in the inferred galaxy properties \citep[e.g.,][]{Conroy_2013ARA&A..51..393C, Diemer_2017ApJ...839...26D, Carnall_2019ApJ...873...44C, Gonzalez_Delgado_2021A&A...649A..79G, Suess_2022ApJ...935..146S, Whitler_2023MNRAS.519.5859W, Pacifici_2023ApJ...944..141P}. To overcome these challenges, nonparametric SFH models have been developed. These flexible models do not impose a specific functional form and are designed to capture the stochastic nature of SFHs \citep[e.g.,][]{Ocvirk_2006MNRAS.365...46O, Leja_2017ApJ...837..170L, Leja_2019ApJ...876....3L, Iyer_2017ApJ...838..127I, Iyer_2019ApJ...879..116I, Q1-SP044}.

A drawback of the nonparametric SFH models is that they are more computationally demanding compared to parametric SFHs and need broader data coverage. Nevertheless, nonparametric models have the potential to accurately reconstruct the complex SFHs of galaxies, while yielding more reliable estimates of their physical properties. Both \citet{Iyer_2018ApJ...866..120I} and \citet{Lower_2020ApJ...904...33L} validated the galaxy properties inferred from nonparametric SFH models by benchmarking them against mock observations derived from semi-analytic models and cosmological simulations. Moreover, \citet{Leja_2019ApJ...876....3L} assessed the effectiveness of nonparametric SFHs using the SED fitting framework {\tt Prospector} \citep{Leja_2017ApJ...837..170L, Johnson_2021ApJS..254...22J}, showing that these models yield more accurate results than their parametric counterparts. In a follow-up study, \citet{Leja_2019ApJ...877..140L} found that galaxies in the 3D-HST catalog \citep{Skelton_2014ApJS..214...24S} appear systematically older and more massive when using nonparametric SFHs instead of parametric models, consistent with theoretical expectations \citep[e.g.,][]{van_Dokkum_2008ApJ...674...29V, Dave_2008MNRAS.385..147D, Dave_2011MNRAS.415...11D, Lilly_2013ApJ...772..119L}. The results from these studies strongly suggest that nonparametric priors have a clear advantage over traditional parametric SFHs, enabling more reliable results. However, a notable challenge still remains in the dependence of SFHs on the choice of physically motivated prior assumptions, as highlighted by \citet{Suess_2022ApJ...935..146S}, \citet{Tacchella_2022ApJ...927..170T}, and \citet{Whitler_2023MNRAS.519.5859W}.

Another potential issue in reconstructing SFHs arises from the modeling and fitting of SEDs from integrated photometry, which is susceptible to the outshining effect \citep{Sawicki_1998AJ....115.1329S, Papovich_2001ApJ...559..620P, Shapley_2001ApJ...562...95S, Daddi_2004ApJ...617..746D, Fontana_2004A&A...424...23F, Pozzetti_2007A&A...474..443P, Trager_2008MNRAS.386..715T, Zibetti_2009MNRAS.400.1181Z, Maraston_2010MNRAS.407..830M, Graves_2010ApJ...717..803G, Pforr_2012MNRAS.422.3285P, Conroy_2013ARA&A..51..393C, Sorba_2015MNRAS.452..235S, Sorba_2018MNRAS.476.1532S, Suess_2022ApJ...935..146S, Whitler_2023MNRAS.519.5859W, Gimenez_Arteaga_2023ApJ...948..126G, Gimenez_Arteaga_2024A&A...686A..63G, Narayanan_2024ApJ...961...73N, Jain_2024MNRAS.527.3291J}. Outshining can occur when a ultraviolet (UV) bright young stellar population (younger than 10~Myr) dominates the light of a galaxy and obscures the light of older stars (older than 200~Myr). This effect can significantly bias the inferred stellar masses and SFHs of galaxies \citep[e.g.,][]{Zibetti_2009MNRAS.400.1181Z, Pforr_2012MNRAS.422.3285P, Sorba_2015MNRAS.452..235S, Jain_2024MNRAS.527.3291J}. While single-aperture spectroscopy can mitigate this bias to some extent, with the inclusion of spectral features, many datasets rely solely on broad- or narrowband photometry.

A more robust alternative involves reconstructing SFHs at spatially resolved scales. Unlike integrated photometry, which captures only global averages, resolved analyses can trace local variations in stellar populations and dust content, enabling more accurate estimates of stellar mass, SFR, and formation timescales, such as the time by which 50\% of a galaxy's stellar mass was formed \citep{Mosleh_2020ApJ...905..170M, Suess_2019ApJ...885L..22S, Abdurrouf_2022ApJ...926...81A, Jain_2024MNRAS.527.3291J, Mosleh_2025ApJ...983..181M}. Studies such as \citet{Gimenez_Arteaga_2023ApJ...948..126G, Gimenez_Arteaga_2024A&A...686A..63G}, and \citet{Jain_2024MNRAS.527.3291J} demonstrate that accounting for spatial structure reduces the impact of outshining, resulting in more accurate stellar mass measurements. The results by \citet{Jain_2024MNRAS.527.3291J} also motivate the use of flexible SFH models on global scales, a result particularly important at higher redshifts, where SFHs are often more bursty \citep{Tacchella_2020MNRAS.497..698T, Narayanan_2024ApJ...961...73N} and thus more prone to outshining effects.

The advent of the \Euclid space telescope \citep{Laureijs11, EuclidSkyOverview} offers an unprecedented view of the local Universe \citep[e.g.,][]{Hunt_2025A&A...697A...9H, Saifollahi_2025A&A...697A..10S, Cuillandre_2025A&A...697A..11C}, providing high spatial resolution and sensitivity in one optical band \citep[\Euclid VIS instrument;][]{EuclidSkyVIS}, and three near-infrared bands \citep[\Euclid NISP instrument;][]{EuclidSkyNISP}. By combining \Euclid observations with ancillary imaging data from ground-based and space observatories, and modeling them through spatially resolved SED fitting, we can thoroughly analyze local galaxies, their SFHs, and their physical properties at kiloparsec and sub-kiloparsec scales.

However, before applying any SED fitting methods to \Euclid observations, we first leverage cosmological hydrodynamical simulations. These simulations serve as powerful tools for exploring the origin and evolution of galaxies, while providing crucial benchmarks for testing and refining SED fitting techniques. In a series of \Euclid papers, we employed two complementary approaches to infer the resolved stellar properties of simulated galaxies at $z=0$ using synthetic \Euclid observations. \citet{EP-Kovacic} applied machine learning techniques to map the distribution of physical parameters within well-resolved simulated galaxies. In addition, \citet{EP-Abdurrouf} tested a pipeline for spatially resolved SED fitting of local galaxies with the {\tt piXedfit} package \citep{Abdurrouf_2021ApJS..254...15A}, exploring different stellar population synthesis (SPS) models and incorporating physically motivated priors, such as mass--age and mass--metallicity relations.

In this study, we also employed SED fitting with the goal of reconstructing SFHs on both global and spatially resolved scales for a sample of 25 TNG50-{\tt SKIRT} Atlas galaxies \citep{Baes_2024A&A...683.181B}. We analyze the SFHs derived using two methods: (1) pixel-by-pixel SED fitting and (2) integrated broadband photometry, employing both parametric and nonparametric models. The goal is to assess the feasibility of reconstructing accurate SFHs from \Euclid-like data. Specifically, we tested whether we can improve upon the standard approach to SFH reconstruction, which typically involves applying either resolved rather than global SED fitting, or nonparametric rather than parametric SFHs. While each of these improvements is commonly applied separately, our approach uniquely combines both, leveraging their joint strengths to provide a more comprehensive and robust analysis of galaxy SFHs. Another key objective of this paper is to demonstrate the added value of the \Euclid\ bands in constraining spatially resolved physical properties. This discussion complements the main analysis by showcasing the direct scientific benefits of incorporating \Euclid\ data, which are highly relevant to the broader goals of the \Euclid\ Collaboration.

This paper is organized as follows. In Sect.~\ref{sc:TNG50_images_and_sample}, we provide a brief overview of the datasets and the methodology adopted to prepare the synthetic images from TNG50-{\tt SKIRT} Atlas for the pixel-by-pixel SED fitting. In Sect.~\ref{sc:sed_fitting}, we describe the construction of our physical model within {\tt Prospector} and provide definitions of the various SFH models used in our analysis. In Sect.~\ref{sc:results}, we compare the inferred physical properties from our fiducial SED fitting-run to the ground truth properties from the simulations. The main analysis of this paper is in Sect.~\ref{sc:discussion}, where we discuss the implications of how the assumed SFH models and the spatial resolution affect the inferred physical properties of the galaxies. We summarize our results in Sect.~\ref{sc:sum_conclusions}.

\section{\label{sc:TNG50_images_and_sample} Simulations, sample selection, and image processing} 

In this section, we present a short overview of the TNG50 cosmological hydrodynamical simulation \citep{Pillepich_2019MNRAS.490.3196P, Nelson_2019MNRAS.490.3234N} and the TNG50-{\tt SKIRT} Atlas (Sect.~\ref{subsc:TNG50}), from which we drew our working sample of galaxies (Sect.~\ref{subsc:sample}). The TNG50-{\tt SKIRT} Atlas provides synthetic images of 1154 galaxies, generated using the radiative transfer code {\tt SKIRT} \citep{Camps_2015A&C.....9...20C, Camps_2020A&C....3100381C}. We also give a brief description of the image processing (Sect.~\ref{subsc:pixedfit}) performed with {\tt piXedfit} \citep{Abdurrouf_2021ApJS..254...15A, Abdurrouf_2022ascl.soft07033A}, to create realistic representations that mimic the observational data.

\subsection{\label{subsc:TNG50} Synthetic images of TNG50 galaxies with {\tt SKIRT}}

To perform the pixel-by-pixel SED fitting analysis, we used galaxies from the TNG50 simulation \citep{Pillepich_2019MNRAS.490.3196P, Nelson_2019MNRAS.490.3234N}, the highest-resolution version of the IllustrisTNG cosmological magnetohydrodynamical suite \citep{Marinacci_2018MNRAS.480.5113M, Naiman_2018MNRAS.477.1206N, Nelson_2018MNRAS.475..624N, Pillepich_2018MNRAS.473.4077P, Springel_2018MNRAS.475..676S}. With a baryonic mass resolution of $8.5 \times 10^4$~M$_\odot$ and spatial resolution of 70--140~pc in star-forming regions, TNG50 combines high resolution with a statistically meaningful galaxy population in a cosmological volume, making it well suited for spatially resolved studies.

A sample of 1154 galaxies from the $z = 0$ snapshot of the TNG50 simulation was defined by \citet{Baes_2024A&A...683.181B}, forming the TNG50-{\tt SKIRT} Atlas database. The database features high-resolution, panchromatic images of these galaxies, generated using the radiative transfer code {\tt SKIRT} \citep{Camps_2015A&C.....9...20C, Camps_2020A&C....3100381C}, a 3D Monte Carlo radiative transfer code designed to model the interaction of the stellar radiation field with dust, in different astrophysical systems. Every galaxy in TNG50 is represented by a collection of stellar and gas particles. The stellar particles serve as the primary source of radiation, and depending on their mass, age, and metallicity, {\tt SKIRT} assigns an SED to each one of them. Stellar particles older than 10~Myr are assigned an SED from the \citet{Bruzual_2003MNRAS.344.1000B} SPS models, whereas stellar particles younger than 10~Myr are assigned an SED from the MAPPINGS III templates for star-forming regions \citep{Groves_2008ApJS..176..438G}.

To account for the effects of diffuse dust (absorption and scattering) on the stellar radiation, a constant dust-to-metal fraction ($f_\mathrm{dust} = 0.2$) was assumed \citep{Trcka_2022MNRAS.516.3728T}, based on observational results in nearby galaxies \citep{De_Vis_2019A&A...623A...5D, Galliano_2021A&A...649A..18G, Zabel_2021MNRAS.502.4723Z}. The dust properties were modeled using The Heterogeneous Evolution Model for Interstellar Solids \citep[THEMIS;][]{Jones_2017A&A...602A..46J} framework. The main goal of the TNG50-{\tt SKIRT} Atlas was to produce synthetic images spanning the UV to near-infrared (NIR) wavelengths. Consequently, the modeling of dust emission was not included in the radiative transfer analysis.

The TNG50-{\tt SKIRT} Atlas originally included synthetic images in 18 wavebands, covering the UV to NIR spectrum (0.15--4.6~$\mu\mathrm{m}$): GALEX~far-UV (FUV) and near-UV (NUV) bands, Johnson~$UBVRI$, LSST~$ugrizy$, 2MASS~$JHK_\mathrm{s}$, and WISE~W1 and W2. \citet{EP-Kovacic} extended this dataset by adding synthetic images for the four \Euclid broadband filters (\IE, \YE, \JE, \HE). Each galaxy in the TNG50-{\tt SKIRT} Atlas was simulated from five different viewing angles, resulting in 22-band images for each projection, with a consistent field of view of 160~kpc~$\times$~160~kpc. These synthetic images are in units of megajanskys per steradian and have a physical pixel size of 100~pc. For the pixel-by-pixel SED fitting analysis, only a subset of these bands is utilized: GALEX FUV and NUV, LSST $u$, $g$, $r$, $i$, $z$, and \Euclid \IE, \YE, \JE, and \HE. Although LSST observations from the Rubin Observatory \citep{Ivezic_2019ApJ...873..111I} are not yet available, other large ground-based campaigns are being carried out to complement \Euclid's observations and can be used to perform spatially resolved SED fitting on nearby galaxies. In particular, the Ultraviolet Near Infrared Optical Northern Survey (UNIONS; Gwyn et al., in prep.) in the Northern Hemisphere provides observations in the $u$, $g$, $r$, $i$, and $z$ bands, while the Dark Energy Survey \citep[DES;][]{Abbott_2021ApJS..255...20A} in the Southern Hemisphere provides observations in the $g$, $r$, $i$, and $z$ bands.

For this work, we used a subset of 11 bands spanning the UV to NIR (GALEX~FUV/NUV, LSST~$ugriz$, and \Euclid \IE, \YE, \JE, and \HE), matching expected multiwavelength coverage from ongoing and upcoming surveys. Alongside synthetic images, the TNG50-{\tt SKIRT} Atlas provides maps of key physical properties such as stellar mass surface density ($\Sigma_\star$), stellar metallicity ($Z_\star$), mass-weighted stellar age ($t_\star$), dust mass surface density maps ($\Sigma_\mathrm{dust}$), and SFR surface density ($\Sigma_\mathrm{SFR}$). These maps serve as ground-truth references for testing our SED fitting.

\subsection{\label{subsc:sample} Sample of simulated galaxies}

\begin{figure}[t]
    \centering
    \includegraphics[width=\columnwidth]{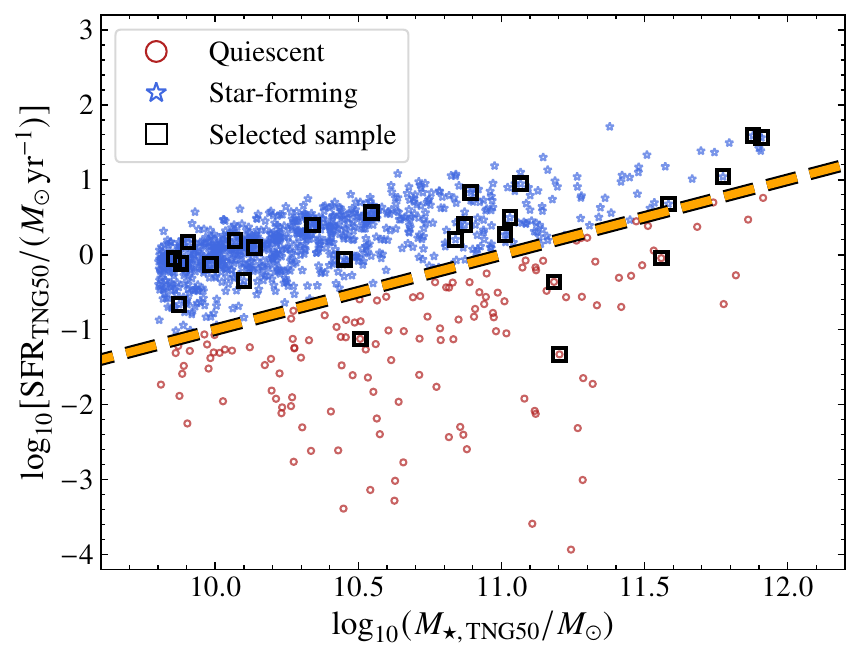}
    \caption{SFR as a function of $M_\star$ for the TNG50-{\tt SKIRT} Atlas galaxies at $z = 0$. Galaxies are separated as star-forming (blue stars) and quiescent (red points) based on their sSFR. The dashed orange line indicates this separation between star formation and quiescence (sSFR = $10^{-11}$~yr$^{-1}$). The black squares indicate the selected sample of 25 galaxies used in our analysis.}
    \label{fig:fig_1}
\end{figure}

\begin{figure*}[t]
    \centering
    \includegraphics[width=\textwidth]{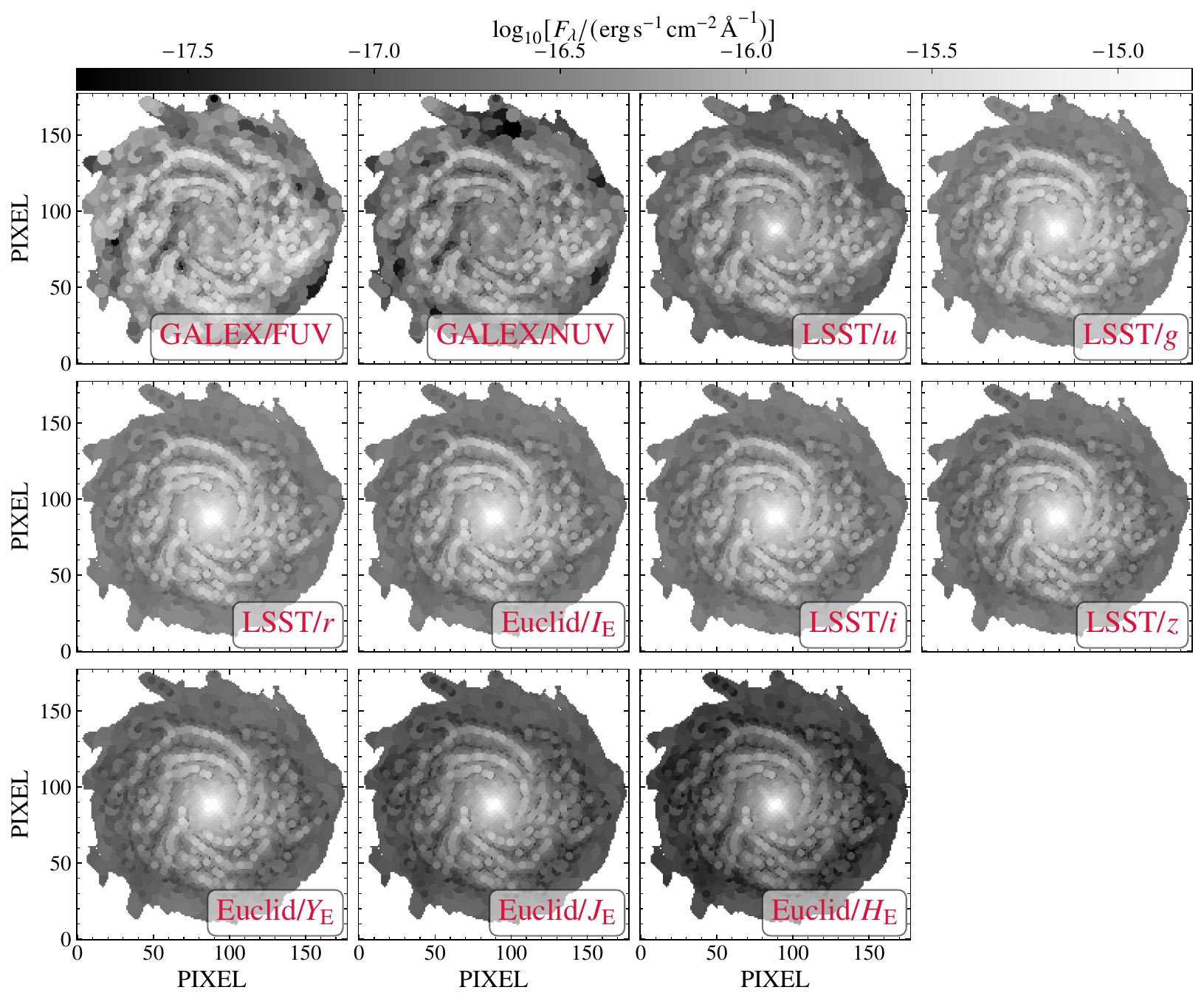}
    \caption{Example of a simulated galaxy processed with {\tt piXedfit}. We show the synthetic images of GALEX, LSST, and \Euclid for TNG275545 with the O1 orientation index. Observational effects were introduced to the original synthetic images from the TNG50-{\tt SKIRT} Atlas database in the form of simulated noise and convolution with the PSF of the corresponding cameras. The FWHM of the PSF as well as the pixel size of the images are $5\arcsecf05$ and $1\arcsecf5$, respectively. The pixels are binned with {\tt piXedfit} to increase the S/N of the spatially resolved SEDs.}
    \label{fig:fig_2}
\end{figure*}

\begin{table*}
\caption{Free parameters and their associated prior distribution functions in the {\tt Prospector} physical model.}
\begin{center}
\scalebox{0.95}{
\begin{threeparttable}
\begin{tabular}{llr}
\hline 
\hline\\
Parameter & Description & Prior distribution \\
\hline\\
$z$ & redshift & fixed at $z = 0.03$\\[0.2cm]
$\logten\left(M_\star/M_\odot\right)$ & total stellar mass formed & uniform: $\rm{min}=5$, $\rm{max}=12$\\[0.2cm]
$Z_\star~\left(\mathrm{Z}_\odot\right)$ & stellar metallicity & uniform: $\rm{min}=0.1$, $\rm{max}=1.55$\\[0.2cm]
SFR ratios & Ratio of adjacent SFRs in the   & Student’s t-distribution with\\
           & eight-bin nonparametric SFH, & $\sigma=0.3$ and $\nu=2$\\
           & seven free parameters in total & \\

  \hline
  & & \\
$\hat{\tau}_{\mathrm{dust}, 2}$ & diffuse dust optical depth & clipped normal: $\rm{min}=0$, $\rm{max}=4$, $\rm{mean}=0.3$, $\sigma=1$\\[0.2cm]
$\hat{\tau}_{\mathrm{dust}, 1}$ & birth-cloud dust optical depth & clipped normal in ($\hat{\tau}_{\mathrm{dust}, 1}/\hat{\tau}_{\mathrm{dust}, 2}$): $\rm{min}=0$, $\rm{max}=4$, $\rm{mean}=0.3$, $\sigma=1$\\[0.2cm]
$n$ & slope of \citet{Kriek_2013ApJ...775L..16K} dust law & uniform: $\rm{min}=-1$, $\rm{max}=0.4$\\
\hline
\end{tabular}
\end{threeparttable}}
\label{tab:table_1}
\end{center}
\end{table*}

The simulated galaxies of TNG50-{\tt SKIRT} Atlas span a wide range of stellar mass ($M_\star$), from $10^{9.8}$~$M_\odot$ to $10^{12}$~$M_\odot$. Figure~\ref{fig:fig_1} shows the star-forming sequence \citep[SFS;][]{Strateva_2001AJ....122.1861S, Baldry_2004ApJ...600..681B, Daddi_2007ApJ...670..156D, Noeske_2007ApJ...660L..43N, Elbaz_2007A&A...468...33E, Whitaker_2012ApJ...754L..29W}, that is the relationship between SFR and $M_\star$, as obtained from the simulations. Galaxies can be separated into two broad galaxy populations: quiescent and star-forming. The two most common criteria to separate these two classes are based on (1) their locus on color-color diagrams \citep[e.g.,][]{Labbe_2005ApJ...624L..81L, Wuyts_2007ApJ...655...51W, Williams_2009ApJ...691.1879W, Arnouts_2013A&A...558A..67A, Schawinski_2014MNRAS.440..889S, Q1-SP031}, and (2) their specific SFR (${\rm sSFR} = \mathrm{SFR}/M_\star$). In this work, we choose to define quiescence based on the sSFR as a more robust method. We define the boundary between quiescent and star-forming galaxies based on sSFR, adopting a threshold of $\logten \left(\mathrm{sSFR} / \mathrm{yr}^{-1} \right) = -11$ \citep{Brinchmann_2004MNRAS.351.1151B, Fontanot_2009MNRAS.397.1776F, Cecchi_2019ApJ...880L..14C, Donnari_2019MNRAS.485.4817D, Paspaliaris_2023A&A...669A..11P, Nersesian_2025A&A...695A..86N}, which yields 285 quiescent and 869 star-forming galaxies.

From the parent sample of TNG50-{\tt SKIRT} Atlas galaxies, a subset of 25 galaxies was selected by \citet{EP-Abdurrouf} for benchmarking spatially resolved SED-fitting. The galaxies were selected based on stellar mass and sSFR criteria, spanning a broad range of sSFRs ($10^{-13}$--$10^{-9}$~yr$^{-1}$) and covering the same stellar mass range as the parent sample, while also representing a diversity of morphological types. Another criterion of the selection process was to prioritize galaxies with an almost face-on orientation, minimizing systematic effects caused by increased dust attenuation in edge-on views. Focusing on this small subsample enables a proof-of-concept analysis of spatially resolved SED fitting with nonparametric SFHs, while keeping the computational demands tractable. Figure~\ref{fig:fig_1} illustrates the location of these 25 galaxies on the SFR--$M_\star$ plane, where 21 are classified as star-forming and 4 as quiescent.

\subsection{\label{subsc:pixedfit} Observational effects and image processing with {\tt piXedfit}}

The synthetic images from the TNG50-{\tt SKIRT} Atlas were post-processed to include realistic observational effects before applying pixel-by-pixel SED fitting. Observational noise (sky background and photon shot noise) was added following prescriptions for the \Euclid Wide Survey and LSST \citep{Martinet-EP4, Merlin-EP25}, as implemented by \citet{EP-Abdurrouf}. The adopted $10\sigma$ ($\mathrm{S}/\mathrm{N}=10$) limiting magnitudes within a $2\arcsec$ aperture for the \IE, \YE, \JE, \HE, and LSST~$u$, $g$, $r$, $i$, and $z$ bands are 24.6, 23.0, 23.0, 23.0, 23.6, 24.5, 23.9, 23.6, and 23.4, respectively \citep[see Table~1 of][]{EP-Abdurrouf}. The limiting magnitudes of the synthetic images quoted here closely match the magnitude limits reported in \Euclid's Quick Data Release \citep[Q1; e.g.,][]{Q1-SP031, Q1-TP005}. For the UV bands, background noise was added based on GALEX MIS observations \citep{Bianchi_2014AdSpR..53..900B}, which provide deeper coverage than AIS and thus stronger constraints on recent star formation.

Each galaxy was placed at $z=0.03$, and images were resampled to the pixel scales of their respective instruments. Following noise addition, all bands were convolved to their instrument point spread functions (PSFs), and subsequently homogenized in spatial resolution and sampling using the {\tt piXedfit} pipeline \citep{Abdurrouf_2021ApJS..254...15A, Abdurrouf_2022ascl.soft07033A}. This procedure ensures that all multiwavelength images share the resolution of the lowest-quality dataset, in this case GALEX, with a full width half maximum (FWHM) of $5\arcsecf05$, and a pixel size of $1\arcsecf5$. Despite the lower resolution, GALEX provides critical UV coverage to constrain recent star formation.

To test the effects of wavelength coverage and variations in imaging datasets, \citet{EP-Abdurrouf} analyzed three sets of imaging data cubes: (1) a combination of GALEX, LSST, and \Euclid (11 bands); (2) LSST and \Euclid (9 bands); and (3) \Euclid-only images. They showed that \Euclid images alone can accurately recover stellar mass surface density. However, to constrain SFHs, shorter-wavelength information is critical. Therefore, for the purposes of our analysis we use the combined images of GALEX, LSST, and \Euclid (11 bands). To improve pixel-level S/N, {\tt piXedfit} applies adaptive binning by grouping neighboring pixels with similar SED shapes, while preserving spatial information. Across the 25 selected galaxies, this yields 6958 pixel bins used in our analysis (see Fig.~\ref{fig:fig_2} for an example). We refer to \citet{EP-Abdurrouf} for further details on the procedure.

\section{\label{sc:sed_fitting} SED fitting with {\tt Prospector} on resolved and global scales}

We applied different physical models to reconstruct the SFHs of the 25 TNG50-{\tt SKIRT} Atlas galaxies, on both global and resolved scales. In this section, we provide a description of those physical models generated with the {\tt Prospector} inference framework \citep{Leja_2017ApJ...837..170L, Johnson_2021ApJS..254...22J}, to fit the photometric SEDs on spatial scales and global scales.

\subsection{\label{subsc:physical_models} Standard physical model within {\tt Prospector}} 

{\tt Prospector} is an SED fitting code that utilizes Bayesian forward modeling and Markov chain Monte Carlo (MCMC) sampling to explore the parameter space. {\tt Prospector} can generate gridless, "on-the-fly" SEDs by combining models of stellar, nebular, and dust components into composite stellar populations. Within {\tt Prospector}, there are numerous methods for treating the SFH of a galaxy. One of {\tt Prospector}'s key strengths is its ability to employ nonparametric SFHs with various prior distributions and parameterizations. 

One of our main goals is to determine which approach returns more meaningful results for the SFH. Since we treat the SFH in different ways, establishing a standard physical model that is consistently applied across all SED fitting runs is essential. Our physical model largely follows the model presented in \citet{Leja_2017ApJ...837..170L, Leja_2018ApJ...854...62L, Leja_2019ApJ...877..140L}. We also used the nested sampler {\tt dynesty} \citep{Skilling_2004AIPC..735..395S, Skilling_2006AIPC..872..321S, Koposov_2022zndo...7388523K} to dynamically sample the parameter space and maximize a chosen objective function as the fit proceeds.

Regarding the spectra of the stellar populations, {\tt Prospector} utilizes the Flexible Stellar Populations Synthesis ({\tt FSPS}) code \citep{Conroy_2009ApJ...699..486C}. We adopted the default SPS parameters in {\tt FSPS}, that is the MILES stellar library \citep{Sanchez_Blazquez_2006MNRAS.371..703S} and the MIST isochrones \citep{Choi_2016ApJ...823..102C}. MILES is an empirical stellar library of high spectral resolution (FWHM~$=2.5$~\AA), with a rest-frame wavelength range from 3525 to 7400~\AA. The FSPS code also covers the UV and NIR wavelength range but with lower spectral resolution. The MIST models are based on the open-source stellar evolution package MESA \citep{Paxton_2011ApJS..192....3P, Paxton_2013ApJS..208....4P, Paxton_2015ApJS..220...15P, Paxton_2018ApJS..234...34P}. \citet{Baes_2024A&A...683.181B} used the \citet[][BC03]{Bruzual_2003MNRAS.344.1000B} SPS models to generate the synthetic images of the TNG50-{\tt SKIRT} Atlas. \citet{EP-Abdurrouf} tested whether a systematic difference exists when using different SPS models (FSPS, BC03\footnote{\citet{EP-Abdurrouf} used the 2016 version of the \citet{Bruzual_2003MNRAS.344.1000B} SPS models.}) with SED fitting and found only a negligible effect on the results. Therefore, the use of FSPS models in this work remains valid. Throughout this paper, we adopted the \citet{Chabrier_2003PASP..115..763C} initial mass function (IMF).  

A common practice in SED fitting is to use a flat prior in logarithmic space when sampling stellar metallicity values \citep{Leja_2017ApJ...837..170L, Leja_2018ApJ...854...62L}. However, this approach tends to favor the exploration of lower metallicity values, which can influence the fitting results. When fitting only photometric data, a broader parameter space needs to be considered, requiring wider steps for stellar metallicities. To mitigate this issue, a flat prior in linear space can be used \citep{Nersesian_2025A&A...695A..86N}. In Appendix~\ref{apdx:A}, we show that a flat prior in linear space yields improved $Z_\star$ estimates in respect to the flat prior in logarithmic space. Thus, in our analysis we adopted a flat prior in linear space for the stellar metallicity. However, as \citet{Nersesian_2024A&A...681A..94N} demonstrated, reliable stellar metallicity constraints remain challenging, even with both broadband and narrowband photometry. Alternatively, employing more physically motivated priors such as a mass--metallicity or age--metallicity relation based on empirical studies of the local Universe \citep[e.g.,][]{Bundy_2015ApJ...798....7B} could potentially improve metallicity estimates \citep[e.g.,][]{Q1-SP044, EP-Abdurrouf}, but it may also introduce biases into the inferred mass--metallicity relationship. 

Since our analysis relies solely on broadband photometry without narrowband or spectroscopic data, and to optimize computational efficiency, we chose not to include nebular emission in our modeling. Finally, the impact of dust grains on the light from stellar populations at UV and optical wavelengths is modeled using a variable dust attenuation law. The strength of the UV bump is constrained according to the findings of \citet{Kriek_2013ApJ...775L..16K}, while the effects of dust attenuation from diffuse dust in the interstellar medium (ISM) and from birth clouds are accounted for separately. The birth-cloud component attenuates the stellar emission from stars with an age up to 10~Myr, and its optical depth is given by

\begin{equation} \label{eq:att_1}
\\\\\ \hat{\tau}_{\mathrm{dust}, 1} = \hat{\tau}_{1} \left(\lambda/5500~\mathrm{\AA}\right)^{-1}.
\end{equation}

\noindent The diffuse dust component has a flexible function that attenuates both stellar and nebular emission of a particular galaxy, using a power-law-modified starburst curve \citep{Calzetti_2000ApJ...533..682C}, extended with the \citet{Leitherer_2002ApJS..140..303L} curve. We use \citep{Noll_2009A&A...507.1793N}

\begin{equation} \label{eq:att_2}
\\\\\ \hat{\tau}_{\mathrm{dust}, 2} = \frac{\hat{\tau}_{2}}{4.05} \left[k^\prime\left(\lambda\right) + D\left(\lambda\right)\right]\left(\frac{\lambda}{\lambda_V}\right)^{n},
\end{equation}

\noindent where $k^\prime\left(\lambda\right)$ is the original \citet{Calzetti_2000ApJ...533..682C} attenuation curve, and $D\left(\lambda\right)$ is a Lorentzian-like Drude profile characterizing the UV bump at 2175~\AA~in the attenuation curve. The normalization of the optical depth of the birth-cloud and diffuse dust components, along with the power-law index ($n$), which describes the shape of the attenuation curve for the diffuse dust component, are set as free parameters in our model \citep[see][for more details]{Leja_2017ApJ...837..170L}.

We should note here that the {\tt SKIRT} radiative transfer calculations produce effective attenuation curves that depend on the underlying dust geometry, composition, and viewing angle. In contrast, {\tt Prospector} assumes a simplified, parametric attenuation law such as the one described above. This choice is intentional, as it mirrors the conditions of real observations where the true dust geometry is unknown and must be approximated through empirical or semiempirical laws. Consequently, part of the dust-age degeneracy discussed in Sect.~\ref{subsc:recovery_of_stellar_props} may be driven by this modeling simplification, reflecting an intrinsic limitation of SED fitting under uncertain dust geometries.

While the TNG50-{\tt SKIRT} Atlas provides dust attenuation maps derived from the radiative transfer simulations, these maps trace only the diffuse dust component. The attenuation associated with birth clouds in star-forming regions is modeled separately in {\tt SKIRT}, making it nontrivial to construct a total, self-consistent attenuation map. Therefore, a direct comparison between the dust attenuation inferred from our SED fitting and that produced by {\tt SKIRT} would be inconsistent and potentially misleading.

\subsection{\label{subsc:sfh_treatments} SFH treatments}

For our fiducial run, we performed a pixel-by-pixel SED fitting, using a flexible nonparametric SFH with a Student's-t prior distribution (also known as the "continuity" prior) described thoroughly in \citet{Leja_2019ApJ...876....3L}. Based on the regularization schemes by \citet{Ocvirk_2006MNRAS.365...46O} and \citet{Tojeiro_2007MNRAS.381.1252T}, the "continuity" prior favors a piecewise constant SFH without sharp transitions in SFR(t). We used eight time elements to describe the nonparametric SFH, specified in look-back time. The recent variations in the SFH were captured by fixing the first two time bins at 0--30~Myr and 30--100~Myr. We also fixed the oldest time bin, placed at $0.85~t_\mathrm{univ} - t_\mathrm{univ}$ where $t_\mathrm{univ}$ is the age of the Universe at the observed redshift ($z = 0.03$). The remaining five bins were spaced equally in logarithmic time between 100~Myr and $0.85 \, t_\mathrm{univ}$. The stellar metallicity was assumed to be constant across all eight time bins. In other words, a single metallicity parameter describes the entire stellar population. Our fiducial model includes 12 free parameters in total. A list of the free parameters and their associated prior distribution functions are given in Table~\ref{tab:table_1}.

In addition to our fiducial model, we performed a pixel-by-pixel SED fitting, using a parametric SFH with a delayed exponential of the following functional form:

\begin{equation} \label{eq:param_SFH}
\\\\\ \mathrm{SFR} \propto t_\mathrm{age} \, {\rm e}^{-t_\mathrm{age}/\tau},
\end{equation}

\noindent where $t_\mathrm{age}$ is the look-back age when SFH began, and $\tau$ is the $e$-folding timescale. In order to account for the uncertainty in the stellar age $t_\mathrm{age}$ and in $\tau$, we followed the methodology described in \citet{Jain_2024MNRAS.527.3291J}. Specifically, we created two datasets of SFHs for each pixel of a particular galaxy: one by varying $t_\mathrm{age}$ while keeping the stellar mass and $\tau$ fixed, and one by varying $\tau$ while keeping the stellar mass and $t_\mathrm{age}$ fixed. To retrieve the total parametric SFH of each pixel, we took the median SFH of the respective sets containing $N$ SFHs, then we calculated the final SFH by averaging the SFHs accounting for the uncertainties in $t_\mathrm{age}$ and $\tau$.

\begin{table}[t]
    \caption{Definitions and nomenclature of the different SFH models in our analysis.}
    \centering
    \scalebox{0.85}{
    \begin{tabular}{l l l}
    \hline
    \hline
    Name &  Scale & Model description\\ 
    \hline
    SFH$_\mathrm{res,\,TNG50}$ & resolved & ground truth SFH from simulations\\
    SFH$_\mathrm{res,\,np}$ & resolved & nonparametric SFH with a "continuity" prior\\
    SFH$_\mathrm{res,\,\tau}$ & resolved & parametric SFH based on a delayed $\tau$-model\\
    SFH$_\mathrm{glob,\,np}$ & global & nonparametric SFH with a "continuity" prior\\
    SFH$_\mathrm{glob,\,\tau}$ & global & parametric SFH based on a delayed $\tau$-model\\
    \hline
    \end{tabular}}
    \label{tab:table_2}
\end{table}

For the purposes of this analysis, we also fit the integrated photometry of our galaxy sample, assuming the same physical model and parameter space given in Table~\ref{tab:table_1}, but again with different treatments of the SFH. In particular, we fit (1) a flexible nonparametric SFH and (2) a delayed $\tau$ model, similar to those SFH models adopted on spatially resolved scales. In Table~\ref{tab:table_2}, we summarize the different SFH models and provide the nomenclature used in our analysis.

Finally, we sought to investigate the impact of the four \Euclid bands on the estimation of the SFHs and stellar properties. For that reason, we performed an additional fitting run using our fiducial SED fitting setup (Table~\ref{tab:table_1}), fitting only the GALEX and LSST bands (seven bands). The results of this analysis are presented in the Appendix~\ref{apdx:B} of this paper. A more comprehensive analysis of how \Euclid observations affect the recovery of galaxy physical properties can be found in \citet{EP-Kovacic} and \citet{EP-Abdurrouf}.

\begin{figure*}[ht!]
    \centering
    \includegraphics[width=\textwidth]{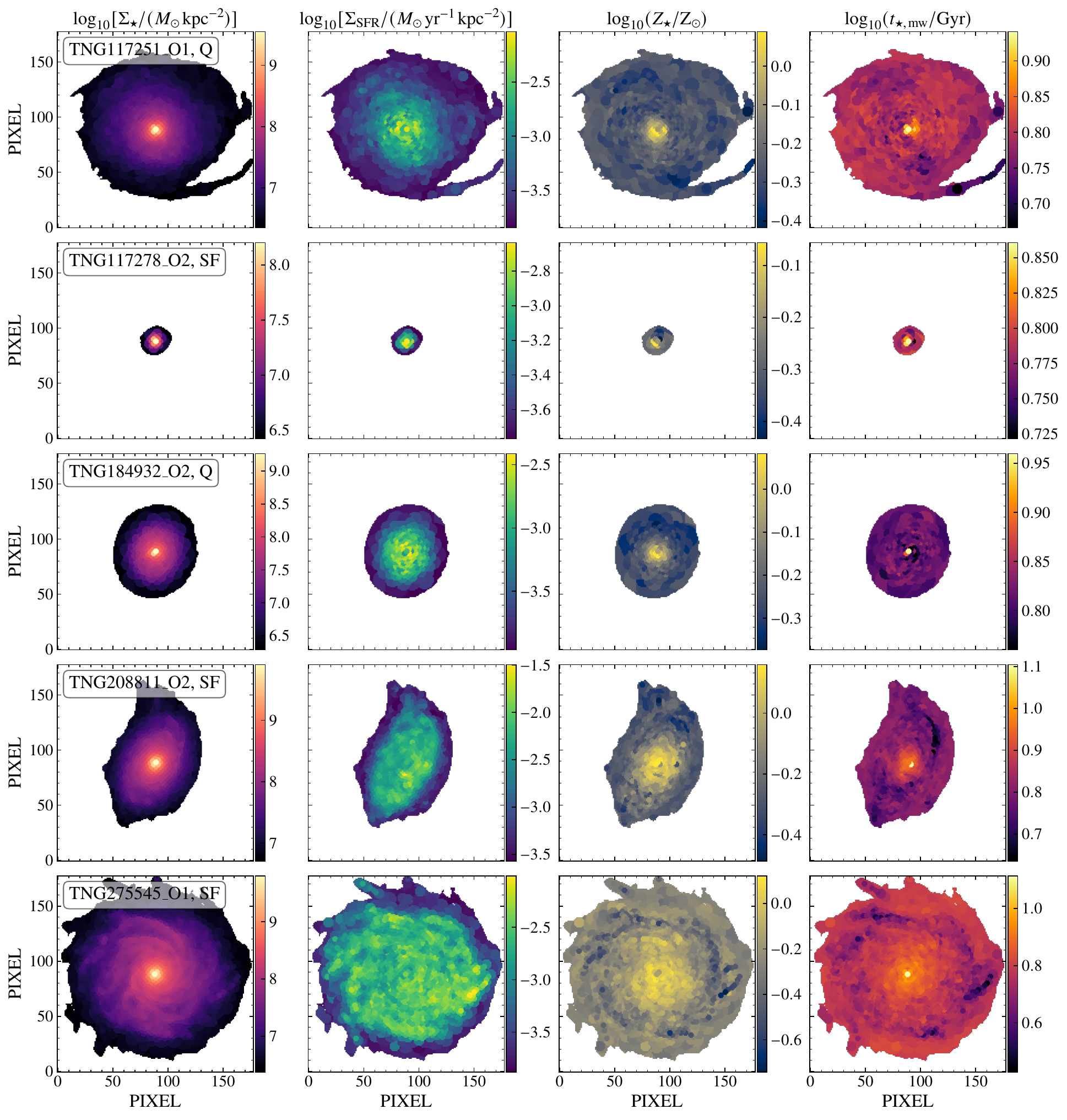}
    \caption{Examples of the stellar property maps of five TNG50-{\tt SKIRT} Atlas galaxies, obtained from the spatially resolved SED fitting of GALEX, LSST, and \Euclid photometry with {\tt Prospector}. The maps include the stellar mass surface density ($\Sigma_\star$), the SFR surface density ($\Sigma_\mathrm{SFR}$), stellar metallicity ($Z_\star$), and mass-weighted stellar age ($t_\mathrm{\star,\,mw}$). Next to each galaxy’s sub-halo ID, we indicate whether a galaxy is classified as star-forming (SF) or quiescent (Q), based on its global sSFR.}
    \label{fig:fig_3}
\end{figure*}

\begin{figure*}[t]
    \centering
    \includegraphics[width=\textwidth]{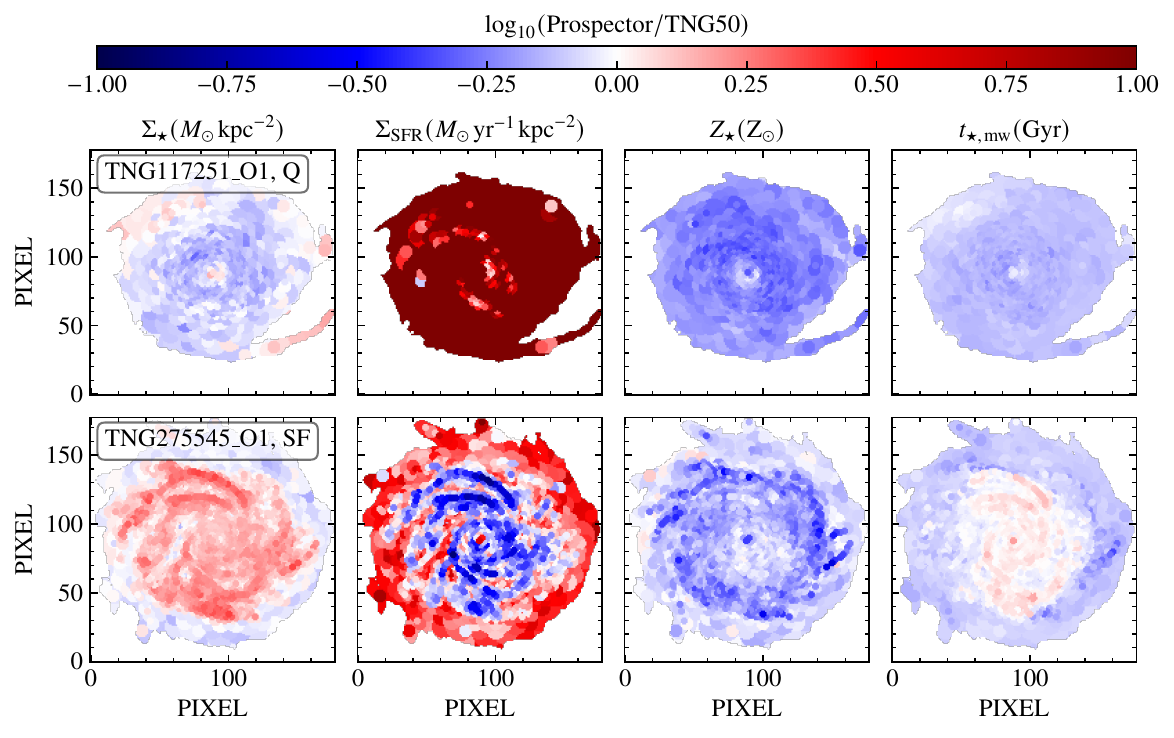}
    \caption{Examples of residual maps of the stellar properties of two TNG50-{\tt SKIRT} Atlas galaxies, obtained from the nonparametric SFH configuration. The residual maps include the stellar mass surface density ($\Sigma_\star$), the SFR surface density ($\Sigma_\mathrm{SFR}$), stellar metallicity ($Z_\star$), and mass-weighted stellar age ($t_\mathrm{\star,\,mw}$). Each residual map is computed as the logarithmic ratio of the inferred property from {\tt Prospector} to the ground truth value from TNG50, expressed as $\logten$({\tt Prospector}/TNG50).}
    \label{fig:fig_4}
\end{figure*}

\section{\label{sc:results} Results}

\subsection{\label{subsc:recovery_of_stellar_props} Recovering the stellar properties at spatially resolved scales}

We present the results from the fiducial spatially resolved SED fitting performed with {\tt Prospector} for the sample of 25 TNG50 galaxies. For each of the 6958 pixel bins analyzed, we adopted the median ($50^\mathrm{th}$ percentile) of the posterior distributions for the inferred physical properties. These values were then used to construct 2D maps of stellar properties, including $\Sigma_\mathrm{SFR}$, $\Sigma_\star$, $Z_\star$, and $t_\mathrm{\star,\,mw}$. Here, the mass-weighted stellar age is defined as the average age of stellar populations, weighted by their mass. In particular, $t_\mathrm{\star,\,mw}$ is given by the following functional form:

\begin{equation} \label{eq:mw_age}
\\\\\ t_\mathrm{\star,\,mw} = \frac{\sum_i M_i \, t_i}{\sum_i M_i},
\end{equation}

\noindent where $M_i$ is the mass of the $i$-th component (e.g., a stellar population of a certain age), and $t_i$ is the age of that component. The denominator is the total stellar mass.

Figure~\ref{fig:fig_3} showcases a few examples of these stellar property maps for the first five galaxies in our sample, sorted by their sub-halo ID. From these maps, we observe that the internal structures of the galaxies are adequately recovered, with detailed resolution of features such as bulges and spiral arms. The results also reproduce well-established trends in the spatial distribution of stellar properties as reported by observational studies \citep[e.g.,][]{Sanchez_Blazquez_2014A&A...570A...6S, Gonzalez_Delgado_2016A&A...590A..44G, Ibarra_Medel_2016MNRAS.463.2799I, Casasola_2017A&A...605A..18C, Garcia_Benito_2017A&A...608A..27G, Zheng_2017MNRAS.465.4572Z, Lopez_Fernandez_2018A&A...615A..27L, Zhuang_2019MNRAS.483.1862Z, Dale_2020AJ....159..195D, Parikh_2021MNRAS.502.5508P, Smith_2022MNRAS.515.3270S, Pessa_2023A&A...673A.147P} and mock observations \citep[e.g.,][]{Nanni_2022MNRAS.515..320N, Nanni_2023MNRAS.522.5479N, Nanni_2024MNRAS.527.6419N}. Specifically, we find that the majority of stellar mass is concentrated in the bulge regions, where stellar populations tend to be older and more metal-rich compared to those in the disk or outskirts. In contrast, $\Sigma_\mathrm{SFR}$ is generally elevated in the spiral arms and outer regions of the galaxies. Inter-arm regions, on average, exhibit lower values of both $\Sigma_\star$ and $\Sigma_\mathrm{SFR}$ compared to the spiral arms \citep[e.g.,][]{Gonzalez_Delgado_2014A&A...562A..47G}.

\begin{figure*}[t]
    \centering
    \includegraphics[width=\textwidth]{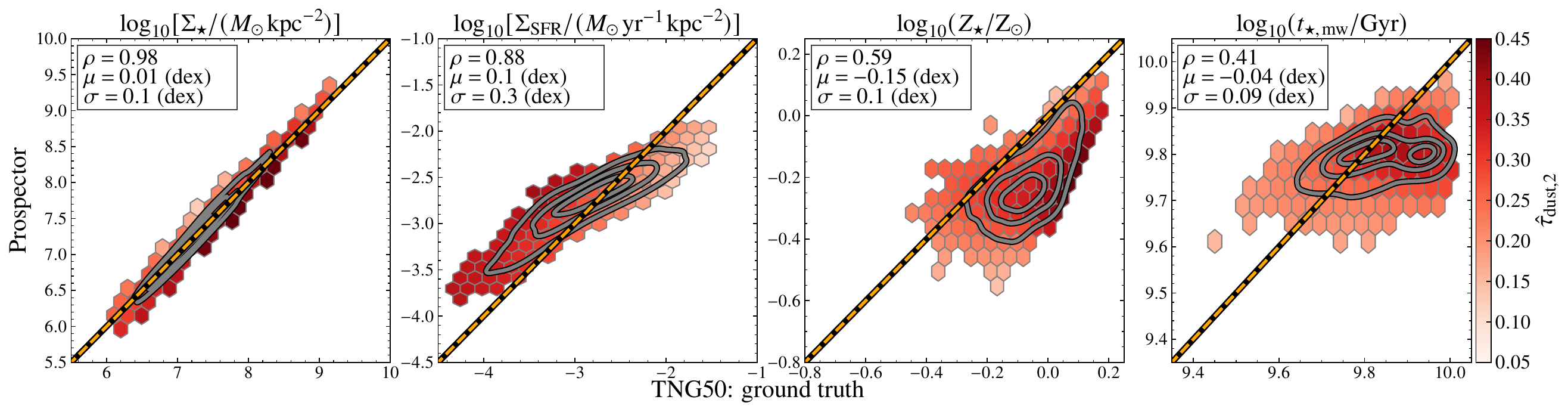}
    \caption{Comparisons of the spatially resolved stellar population properties for 6958 pixel bins in the 25 sample galaxies derived from spatially resolved SED fitting with a nonparametric SFH model. The depicted physical properties include the stellar mass surface density ($\Sigma_\star$), the SFR surface density ($\Sigma_\mathrm{SFR}$), the stellar metallicity ($Z_\star$), and the mass-weighted stellar age ($t_\mathrm{\star,\,mw}$). The colors indicate the average trend with the dust optical depth in the $V$ band ($\hat{\tau}_\mathrm{dust,\, 2}$), calculated with the {\tt LOESS} method. Each hexbin includes a minimum of five points. Contours enclose 20\%, 50\%, and 80\% of the total data. The dashed orange line represents the one-to-one relation. The Spearman’s correlation coefficient ($\rho$), measured bias (average offset $\mu$), and scatter (standard deviation $\sigma$) relative to the ground truth from TNG50 are indicated in the legend of each panel.}
    \label{fig:fig_5}
\end{figure*}

\begin{figure*}[t]
    \centering
    \includegraphics[width=\textwidth]{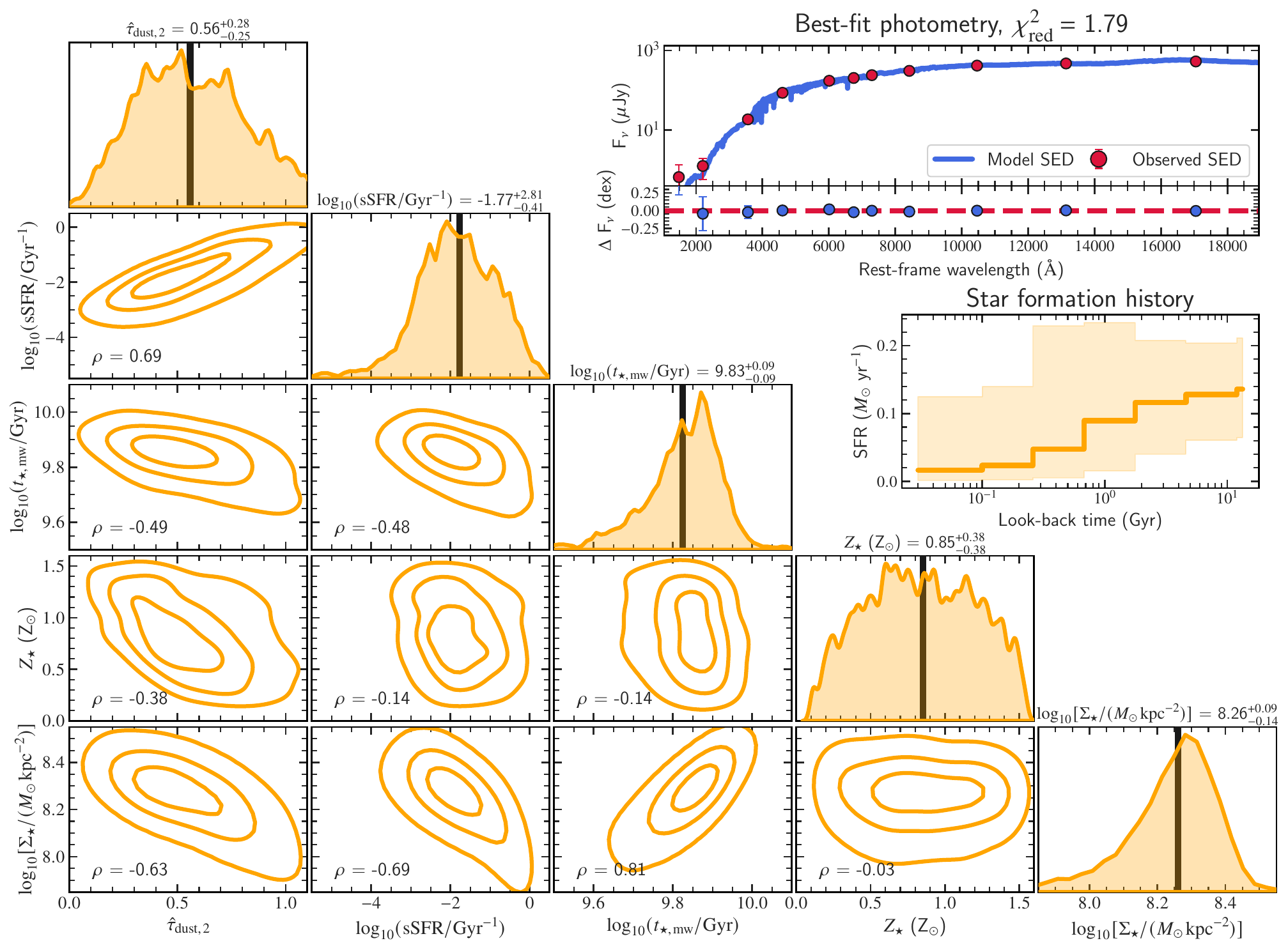}
    \caption{Joint posterior distributions of the main physical properties in our analysis, the best-fit SED, and the recovered SFH for a selected pixel bin of TNG414917 with the O4 orientation index. The corner plot displays the posterior distributions of $\hat{\tau}_\mathrm{ dust,\, 2}$, sSFR, $t_\mathrm{\star,\, mw}$, $Z_\star$, and $\Sigma_\star$. Contours enclose the 20\%, 50\%, and 80\% of the total data, while the vertical dashed black lines mark the median values for each parameter. The median and the $16^\mathrm{th}$--$84^\mathrm{th}$ percentile range for each physical property are given at the top of the corresponding posterior distribution. The Spearman’s correlation coefficient, $\rho$, quantifies the strength of correlation between two posterior distributions. Top right: Comparison of the input photometric data (red points) to the best-fit SED model (blue line). Bottom right: Subpanel illustrating the absolute residuals defined as $\Delta F_{\nu} = \logten(\mathrm{fit}/\mathrm{ground\ truth})$. The photometric uncertainties are not visible in the plot because they are smaller than the symbol size. Bottom: Posterior distribution of the SFH.}
    \label{fig:fig_6}
\end{figure*}

To assess the quality of our fits and evaluate the ability of {\tt Prospector} to recover stellar properties from broadband photometry, we compared our spatially resolved SED fitting results with the 2D maps of ground truth stellar properties from TNG50, generated using {\tt SKIRT}. In Fig.~\ref{fig:fig_4}, we show the residual maps of stellar properties for two example galaxies: a quiescent and a star-forming one. In both cases, the residuals in $\Sigma_\star$ remain within 0.15~dex. For the star-forming galaxy, $\Sigma_\star$ is slightly underestimated in the spiral arms. In the quiescent galaxy, the small number of star-forming gas particles led to only a few usable pixel bins in the $\Sigma_\mathrm{SFR}$ map, with an overall underestimated SFR. For the star-forming galaxy, we observe a general pattern: $\Sigma_\mathrm{SFR}$ is underestimated in the outer disk and overestimated in the inner spiral arms. These residual patterns likely arise from degeneracies between dust attenuation, stellar age, SFH, and metallicity (see also Fig.~\ref{fig:fig_6}). For $Z_\star$, we find a systematic underestimation across both galaxies. Similarly, $t_\mathrm{\star,\,mw}$ is underestimated in the quiescent galaxy. In contrast, for the star-forming galaxy, the ages are well recovered in the inner disk but tend to be underestimated toward the outer edges.

In Fig.~\ref{fig:fig_5}, we present the comparison of the four stellar properties for all spatial pixel bins across the 25 galaxies in our sample. We color-coded the scatter plots by the diffuse dust optical depth ($\hat{\tau}_\mathrm{dust,\, 2}$) in the $V$ band. It is well known that SED fitting with photometry alone can suffer from the age-dust-metallicity degeneracy, as highlighted by previous studies \citep[e.g.,][]{Worthey_1994ApJS...94..687W, Silva_1998ApJ...509..103S, Devriendt_1999A&A...350..381D, Pozzetti_2000MNRAS.317L..17P, Bell_2001ApJ...550..212B, Walcher_2011Ap&SS.331....1W, Diaz_Garcia_2015A&A...582A..14D}. We derived the average trend of $\hat{\tau}_\mathrm{dust,\, 2}$ with the stellar properties with the Locally Weighted Regression ({\tt LOESS}) method \citep{Cleveland_doi:10.1080/01621459.1988.10478639} as implemented in the {\tt LOESS} routine\footnote{https://pypi.org/project/loess/} by \citet{Cappellari_2013MNRAS.432.1862C}. To facilitate quantitative comparisons, we computed the mean offset ($\mu$) and scatter (standard deviation, $\sigma$) for the logarithmic ratio between the best-fit parameters and the ground truth. We also calculated the Spearman rank-order correlation coefficient ($\rho$), indicative of the significance of the relationship between two datasets. The results for each stellar property are summarized at the top-left of each panel in Fig.~\ref{fig:fig_5}.

For $\Sigma_\star$, we find excellent agreement with the ground truth, as evidenced by a small offset ($\mu = 0.01$~dex), low scatter ($\sigma = 0.1$~dex), and a high correlation ($\rho = 0.98$). This suggests that stellar mass can be reliably recovered on spatially resolved scales using photometry, which covers the rest-frame NIR. In particular, this demonstrates the strong potential of \Euclid observations for mapping $\Sigma_\star$ in local galaxies.

For $\Sigma_\mathrm{SFR}$, the recovery is similarly strong ($\rho = 0.88$), but with an increased mean offset ($\mu = 0.1$~dex), and scatter around the mean ($\sigma = 0.3$~dex). In particular, the offset is stronger at the lower end of the distribution, where values of $\Sigma_\mathrm{SFR}$ are very low and comparisons become less meaningful. At these lower values, dust attenuation appears to increase the inferred SFR due to the dust-age degeneracy. At the high end of the distribution, we notice that the $\Sigma_\mathrm{SFR}$ values derived using {\tt Prospector} tend to be lower than the ground truth. These discrepancies in the $\Sigma_\mathrm{SFR}$ maps may stem from the limited number of stellar particles in the latest time bin of the TNG50 $\Sigma_\mathrm{SFR}$ maps, which can result in spurious SFR values.

For $Z_\star$, we measure a significant systematic offset of $\mu = -0.15$~dex, with a low scatter ($\sigma = 0.1$~dex) and a moderate correlation ($\rho = 0.59$). This is an expected result, simply because there is not enough information in broadband photometry to provide meaningful constraints on stellar metallicity \citep{Nersesian_2024A&A...681A..94N, Csizi_2024A&A...689A..37C}. \citet{EP-Abdurrouf} demonstrated that a better recovery of $Z_\star$ can be achieved when using mass-metallicity and age-metallicity priors, highlighting the importance of these prior functions in improving SED fitting with Bayesian methods.

The recovery of the stellar ages is limited due to a narrow dynamic range, with a smaller systematic offset ($\mu = -0.04$~dex), and scatter ($\sigma = 0.09$~dex), yet with a more moderate correlation coefficient ($\rho = 0.41$). Notably, the stellar age estimates show a closer alignment with the ground truth than for $Z_\star$. The largest deviations from the true values occur for older stellar populations, while younger populations are better recovered. These findings are consistent with those by \citet{EP-Abdurrouf}, where more informed priors were used. This suggests that the use of specialized priors may not significantly improve age estimates for this dataset, which covers only the rest-frame UV-optical-NIR regime and lacks optical spectroscopy, an important constraint for age determination \citep{Nersesian_2024A&A...681A..94N, Nersesian_2025A&A...695A..86N}. The absence of spectroscopy, and particularly the lack of spectral features that are key age indicators, such as H$\delta$, likely contributes to this limitation. Finally, color-coding by $\hat{\tau}_\mathrm{dust,\, 2}$ reveals that the secondary peak in stellar ages is linked to the dust-age degeneracy, where {\tt Prospector} infers younger ages, and the red colors are primarily attributed to dust attenuation rather than the presence of older stellar populations.

To investigate these degeneracies, we examined a representative pixel bin with an overestimated SFR and underestimated stellar age and inspected the joint posterior distributions of its inferred properties (Fig.~\ref{fig:fig_6}). The corner plot highlights several well-known correlations: a strong positive correlation between $\hat{\tau}_\mathrm{dust,\, 2}$ and sSFR ($\rho = 0.69$), and moderate anticorrelations of $\hat{\tau}_\mathrm{dust,\, 2}$ with $t_\mathrm{\star,\, mw}$ ($\rho = -0.49$) and $Z_\star$ ($\rho = -0.38$). In addition, $\Sigma_\star$ correlates tightly with $t_\mathrm{\star,\, mw}$ ($\rho = 0.81$), while strongly anticorrelating with both sSFR ($\rho = -0.69$) and dust attenuation ($\rho = -0.63$). These results confirm that the well-known degeneracies between dust, age, star formation, and metallicity persist in the analysis, contributing to the observed biases in the recovered stellar properties. Breaking these degeneracies remains a challenge and likely requires the inclusion of spectroscopic information as well as information in the infrared wavelength regime.

\begin{figure*}[t]
    \centering
    \includegraphics[width=\textwidth]{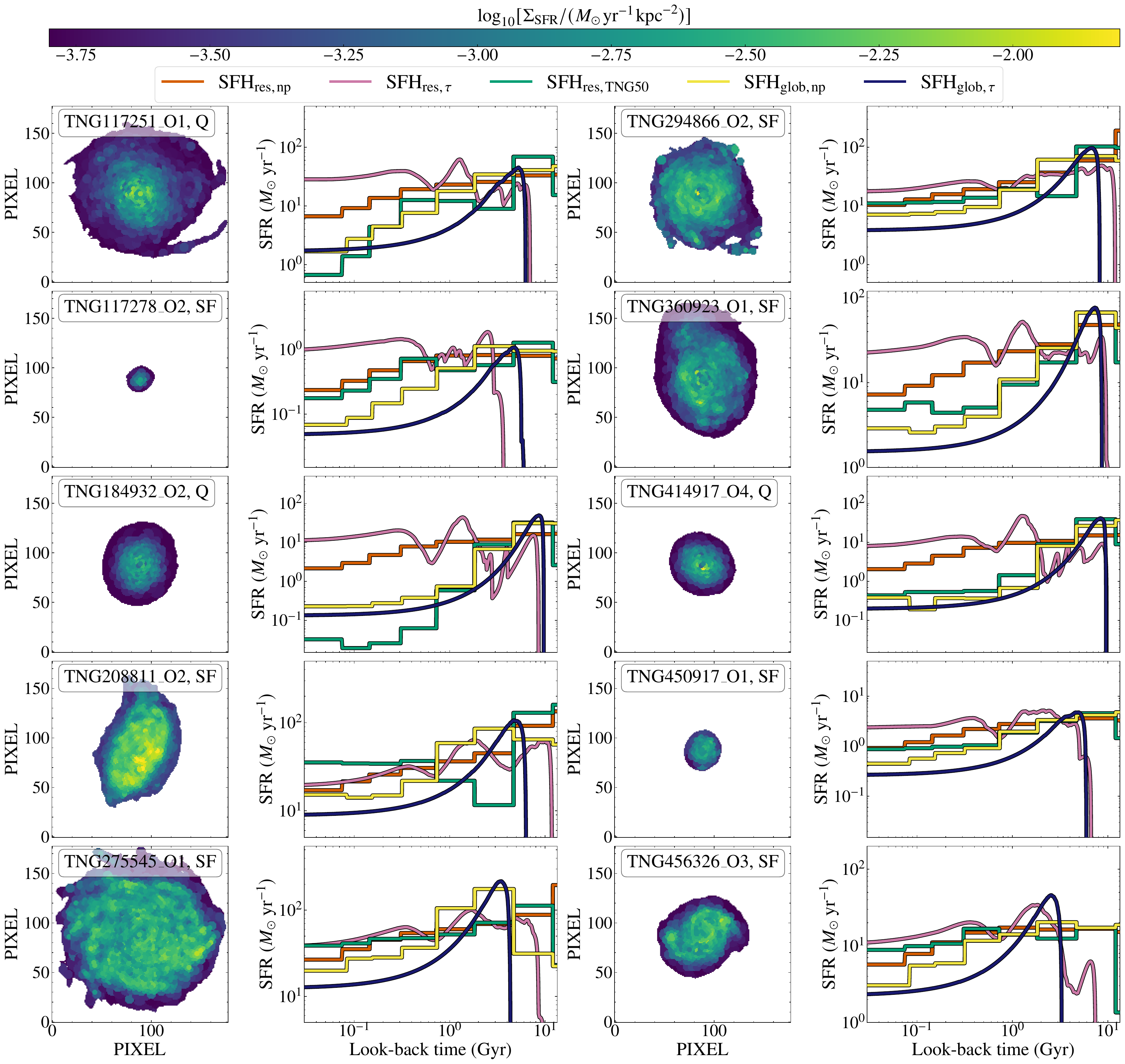}
    \caption{Examples of SFR surface density ($\Sigma_\mathrm{SFR}$) maps and SFH models for ten galaxies in our TNG50-{\tt SKIRT} Atlas sample. Left panel: $\Sigma_\mathrm{SFR}$ map for each galaxy, as estimated with {\tt Prospector}. Right panel: Comparison of various global SFH models to the ground truth SFH from TNG50. Specifically, we show the global nonparametric SFH inferred from our fiducial spatially resolved run (SFH$_\mathrm{res,\,np}$; orange), the global parametric SFH inferred from the spatially resolved map (SFH$_\mathrm{res,\,\tau}$; pink), the nonparametric SFH derived from fitting the integrated photometry (SFH$_\mathrm{glob,\,np}$; yellow), the global parametric SFH based on a simple $\tau$-model (SFH$_\mathrm{glob,\,\tau}$; blue), and the true global SFH from the TNG50 resolved map (SFH$_\mathrm{res,\,TNG50}$; green)}.
    \label{fig:fig_7}
\end{figure*}

\subsection{\label{subsc:SFHs_shapes} Recovering the shapes of galaxy SFHs}

In this section, we present the SFHs of our sample of TNG50 galaxies and evaluate the different approaches for reconstructing the SFH of a galaxy. We analyze the differences between the SFHs obtained from our fiducial spatially resolved SED fitting with {\tt Prospector} (using a nonparametric model) to those derived from resolved fits assuming a parametric SFH, as well as to global-scale fits using both nonparametric and parametric models. These are further compared to the ground truth SFHs from TNG50 on resolved scales. This comparison allows us to assess how modeling assumptions and spatial resolution impact the inferred SFHs. To compute the total SFH of each galaxy from the spatially resolved data, we sum the SFRs across all pixel bins associated with that galaxy, at the specific time bin. This procedure is applied consistently to the spatially resolved SED fitting results, and the TNG50 spatially resolved SFHs.

\begin{figure*}[t]
    \centering
    \includegraphics[width=\textwidth]{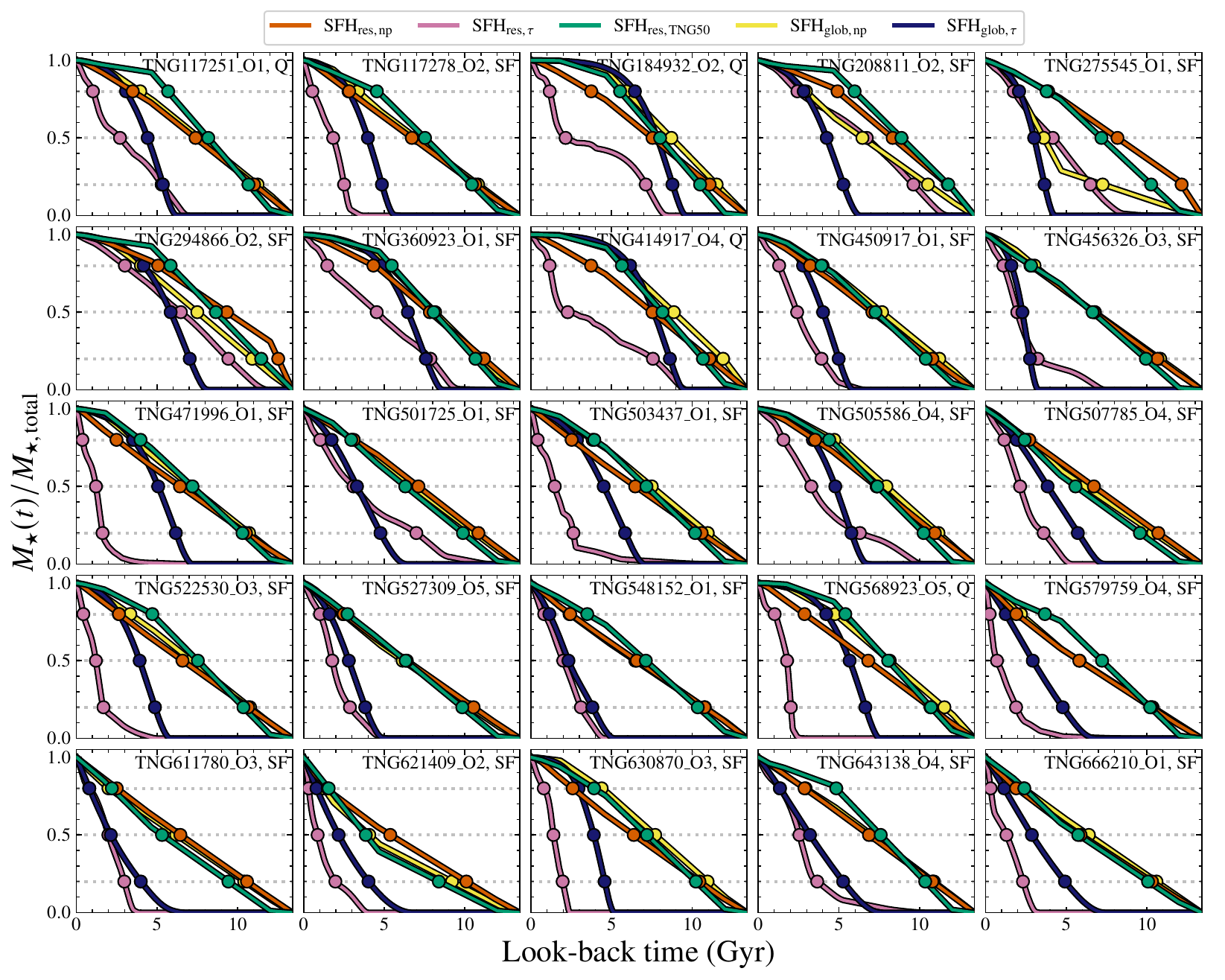}
    \caption{Comparison of the normalized cumulative stellar mass growth curves of all 25 galaxies in our sample. We plot the cumulative stellar mass growth curves from the global nonparametric SFH inferred from our fiducial spatially resolved run (SFH$_\mathrm{res,\,np}$; orange), the global parametric SFH inferred from the spatially resolved map (SFH$_\mathrm{res,\,\tau}$; pink), the nonparametric SFH derived from fitting the integrated photometry (SFH$_\mathrm{glob,\,np}$; yellow), the global parametric SFH based on a simple $\tau$-model (SFH$_\mathrm{glob,\,\tau}$; blue), and the true global SFH from the TNG50 resolved map (SFH$_\mathrm{res,\,TNG50}$; green). The dotted lines indicate the different formation times by which 20\%, 50\%, and 80\% of a galaxy’s stellar mass was formed. Both nonparametric SFHs (SFH$_\mathrm{res,\,np}$ and SFH$_\mathrm{glob,\,np}$) seem to be in better agreement with SFH$_\mathrm{res,\,TNG50}$, while most of the parametric SFHs (both SFH$_\mathrm{res,\,\tau}$ and SFH$_\mathrm{glob,\,\tau}$) are skewed toward a later cosmic time, thus lacking the contribution from the older stellar populations.}
    \label{fig:fig_8}
\end{figure*}

\begin{figure*}[h!]
    \centering
    \includegraphics[width=\textwidth]{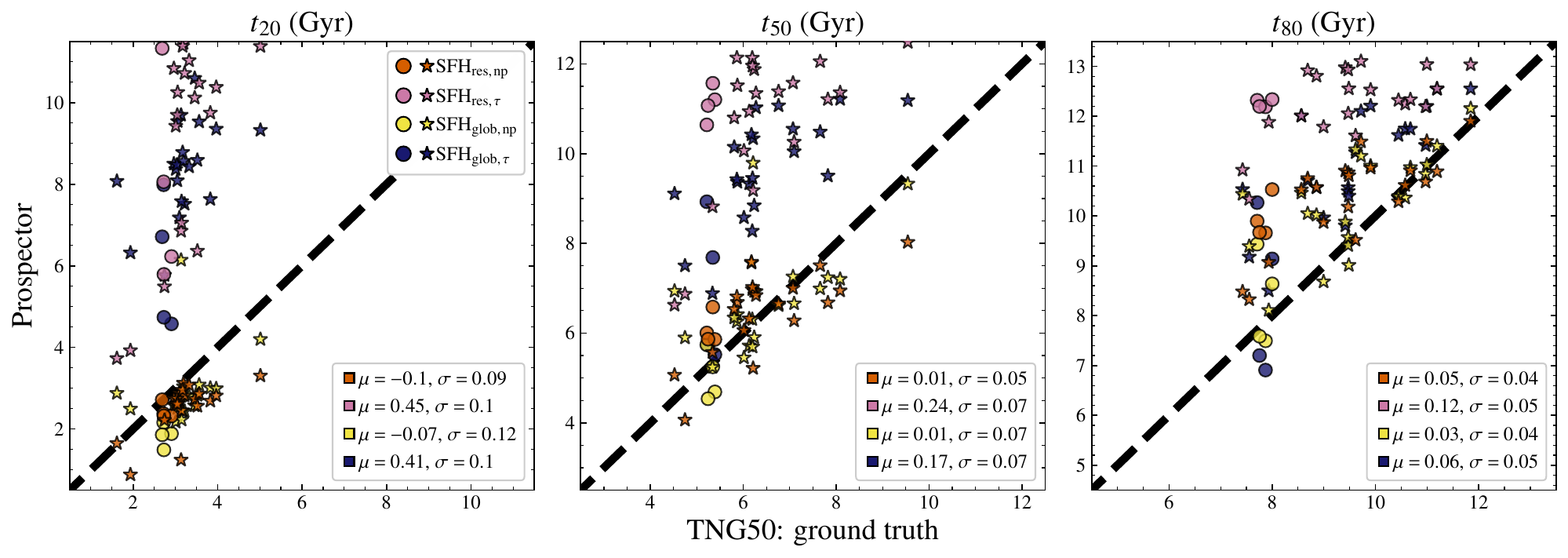}
    \caption{Comparison of the formation timescales for the four different SFH models with respect to the ground truth values from the TNG50 simulation. Each marker represents the formation time of a galaxy in our sample, color-coded according to the corresponding SFH model. Star-forming galaxies are indicated with stars, while quiescent galaxies with points. From left to right: Time for 20\% ($t_{20}$), 50\% ($t_{50}$), and 80\% ($t_{80}$) of a galaxy’s stellar mass formation. The four models include the spatially resolved model from pixel-by-pixel SED fitting with a nonparametric (orange) and parametric (pink) SFH, the integrated SED fitting with a nonparametric SFH (yellow), and a simple $\tau$ model within {\tt Prospector} (blue). The measured bias (average offset $\mu$) and scatter (standard deviation $\sigma$) relative to the ground truth from TNG50 are indicated in the legends of the panels. The formation times inferred from the spatially resolved and integrated nonparametric SFHs are in best agreement with the ground truth values.}
    \label{fig:fig_9}
\end{figure*}

Figure~\ref{fig:fig_7} presents the $\Sigma_\mathrm{SFR}$ maps alongside a comparison of the five global SFH models for the first ten galaxies in our sample, sorted by their sub-halo ID. For each galaxy, the left panel shows the $\Sigma_\mathrm{SFR}$ map derived using the nonparametric SFH model, highlighting the spatial distribution of the present SFR. The right panel compares the four global SFH models reconstructed with {\tt Prospector} to the ground truth SFH from TNG50. The definitions and nomenclature of the different SFH models are given in Table~\ref{tab:table_2}.

A first qualitative inspection of the SFHs shown in Fig.~\ref{fig:fig_7} reveals that the SFH$_\mathrm{res,\,np}$ and SFH$_\mathrm{glob,\,np}$ align more closely with the ground truth SFH$_\mathrm{res,\,TNG50}$ than the parametric SFH$_\mathrm{glob\,\tau}$ and SFH$_\mathrm{res\,\tau}$. While all four SFH models reproduce in general the shape of SFH$_\mathrm{res,\,TNG50}$, the nonparametric models seem to describe better the extended period of star formation. On the other hand, both parametric SFH models fail to match the SFR amplitude of SFH$_\mathrm{res,\,TNG50}$, while the peak of SFR is skewed at a later cosmic time. This mismatch can lead to underestimated stellar masses and younger inferred stellar ages, highlighting the limitations of relying on parametric SFH assumptions.

A more quantitative way to compare SFH models with different parameterization schemes is through an analysis of their cumulative stellar mass growth and formation times. This approach is particularly effective for evaluating how the various SFH models assemble the stellar mass of galaxies over time. Figure~\ref{fig:fig_8} presents the recovered cumulative stellar mass growth curves, normalized to the total stellar mass estimated by each method,\footnote{The bias in the recovered stellar mass for each method is discussed in Sect.~\ref{subsec:bias_in_glob_props}.} as a function of look-back time for the full sample of 25 TNG50-{\tt SKIRT} Atlas galaxies. It is evident that SFH$_\mathrm{res,\,np}$ aligns with the ground truth SFH$_\mathrm{res,\,TNG50}$, capturing both the earlier formation time and the extended duration of star formation. Similarly, SFH$_\mathrm{glob,\,np}$ is in very good agreement with both the ground truth SFH$_\mathrm{res,\,TNG50}$ and the resolved SFH$_\mathrm{res,\,np}$. In contrast, the parametric model, in both resolved (SFH$_\mathrm{res,\,\tau}$) and integrated scales (SFH$_\mathrm{glob,\,\tau}$), shows quite large differences in respect to SFH$_\mathrm{res,\,TNG50}$, with delayed and less extended SFHs.

\begin{table*}[h]
    \caption{Mean offset and standard deviation of formation times and global properties from the TNG50 reference for each SFH model.}
    \centering
    \scalebox{1.2}{
    \begin{tabular}{l rrrrrr}
    \hline
    \hline
    Model & $\Delta \logten (t_{20})$ & $\Delta \logten (t_{50})$ & $\Delta \logten (t_{80})$ & $\Delta \logten (M_\star)$ & $\Delta \logten (\mathrm{sSFR})$ & $\Delta \logten (t_\mathrm{\star,\,mw})$\\ 
          & (Gyr) & (Gyr) & (Gyr) & ($M_\odot$) & (yr$^{-1}$) & (Gyr) \\
    \hline
    SFH$_\mathrm{res,\,np}$    & $-0.10\pm 0.09$ & $0.01\pm 0.05$ & $0.05\pm 0.04$ & $-0.03\pm 0.06$ & $ 0.01\pm 0.19$ & $ 0.01\pm 0.07$ \\
    SFH$_\mathrm{res,\,\tau}$  & $ 0.45\pm 0.10$ & $0.24\pm 0.07$ & $0.12\pm 0.05$ & $-0.12\pm 0.06$ & $ 0.40\pm 0.35$ & $-0.27\pm 0.10$ \\
    SFH$_\mathrm{glob,\,np}$   & $-0.07\pm 0.12$ & $0.01\pm 0.07$ & $0.03\pm 0.04$ & $ 0.01\pm 0.06$ & $-0.17\pm 0.32$ & $ 0.02\pm 0.08$ \\
    SFH$_\mathrm{glob,\,\tau}$ & $ 0.41\pm 0.10$ & $0.17\pm 0.07$ & $0.06\pm 0.05$ & $-0.06\pm 0.06$ & $-0.24\pm 0.34$ & $-0.23\pm 0.11$ \\
    \hline
    \end{tabular}}
    \label{tab:table_3}
\end{table*}

One key observation is that, for most galaxies, the cumulative stellar mass growth predicted by the parametric SFHs is systematically biased toward later formation times. To further quantify these discrepancies, Fig.~\ref{fig:fig_9} compares the formation times at which 20\% ($t_{20}$), 50\% ($t_{50}$), and 80\% ($t_{80}$) of a galaxy’s stellar mass was assembled across the different SFH models. We quantify the differences by calculating the mean offset ($\mu$) and scatter (standard deviation, $\sigma$) relative to the ground truth SFHs from TNG50. The values of $\mu$ and $\sigma$ for each formation time are given in Table~\ref{tab:table_3}.

The nonparametric SFHs, both spatially resolved and global, exhibit strong agreement with the ground truth across all timescales. Our fiducial model reproduces $t_{20}$ within 0.1~dex, while the global nonparametric model (SFH$_\mathrm{glob,\,np}$) shows an even smaller offset of 0.07~dex. Interestingly, both models tend to recover slightly earlier $t_{20}$ values than the reference SFH$_\mathrm{res,\,TNG50}$. In contrast, the parametric SFH models significantly delay the early phases of star formation, and thus underestimate the contribution of the older stellar population.

For $t_{50}$, both nonparametric SFH models achieve an excellent agreement with the ground truth ($\mu = 0.01$~dex), and even for $t_{80}$, the offsets remain small (within $0.05$~dex). These results highlight that spatially resolved fitting with a nonparametric SFH prior can yield accurate reconstructions of galaxy SFHs. Furthermore, we report that nonparametric SFHs at global scales still perform remarkably well. This result suggests that the use of flexible SFHs at global scales can return meaningful formation histories and are less susceptible to the outshining effect compared to parametric SFHs. This provides a significant advantage, especially when spatially resolved data are not available. Crucially, this remains true even in the absence of spectroscopic information, underscoring the power of spatially resolved, and most importantly nonparametric approaches in SFH reconstruction.

\begin{figure*}[t]
    \centering
    \includegraphics[width=\textwidth]{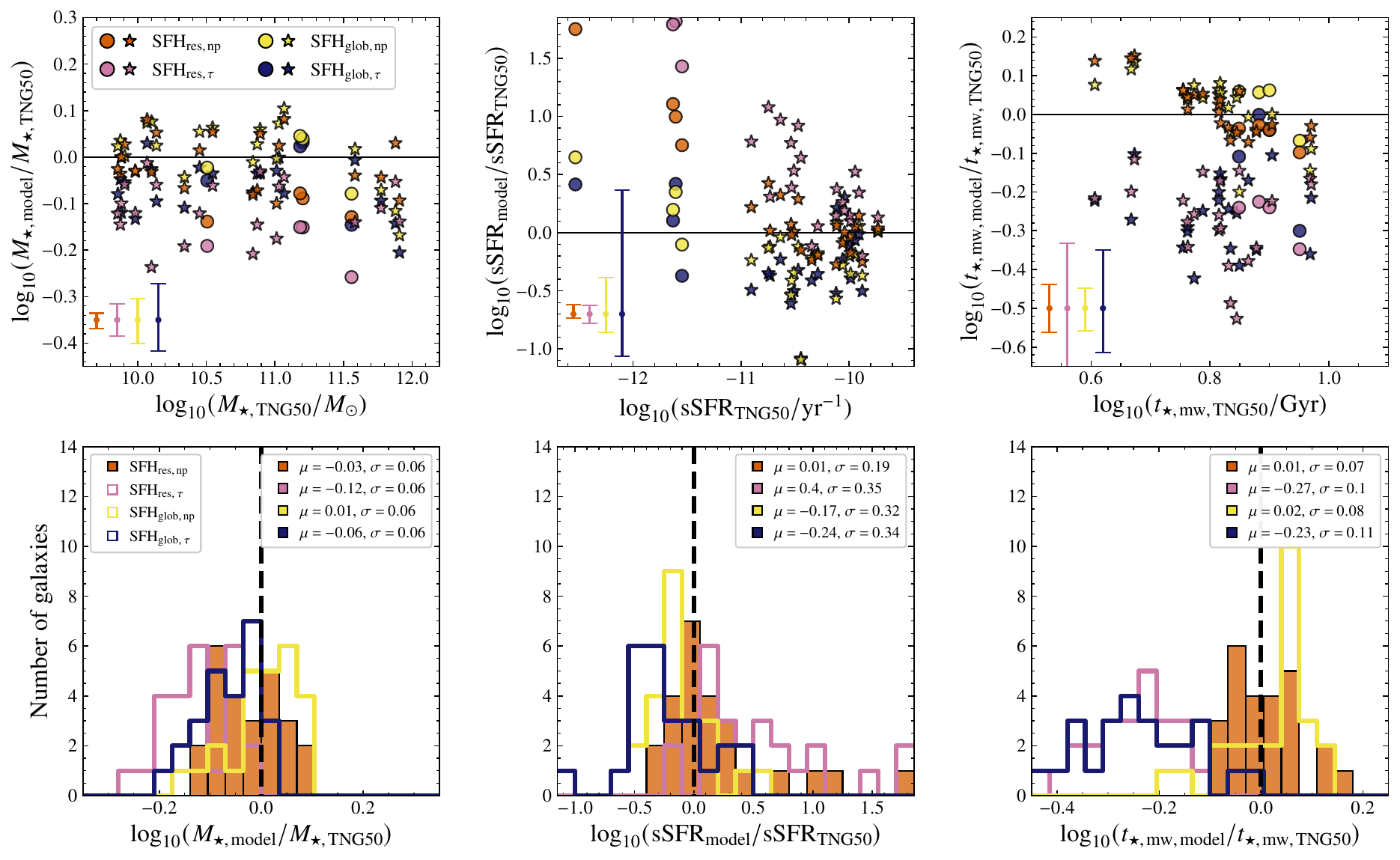}
    \caption{Differences in the global stellar properties for the four different SFH models with respect to the ground truth global values from the TNG50 simulation. From left to right: $M_\star$, sSFR, and $t_\mathrm{\star,\,mw}$. The properties were derived through SED fitting under four different configurations: spatially resolved fitting with a nonparametric SFH$_\mathrm{res,\,np}$ (orange), spatially resolved SED fitting with a simple $\tau$-model SFH$_\mathrm{res,\,\tau}$ (pink), global fitting with a nonparametric SFH$_\mathrm{glob,\,np}$ (yellow), and a global fitting with a simple $\tau$-model SFH$_\mathrm{glob,\,\tau}$ (blue). Typical uncertainties for each configuration are shown in the bottom-left corner of each panel. Star-forming galaxies are indicated with stars, while quiescent galaxies with points. The bottom panels show the residual distributions of each stellar property. The measured bias (average offset) and scatter (standard deviation) are indicated in the legends of the panels. The properties inferred from the spatially resolved and integrated nonparametric SFHs are in best agreement with the ground truth values.}
    \label{fig:fig_10}
\end{figure*}

\subsection{Measuring the bias in global properties}\label{subsec:bias_in_glob_props}

The results presented in this paper suggest that nonparametric SFHs, either at spatially resolved or integrated scales, describe the variation of star formation across cosmic time more accurately. On the other hand, the use of a delayed $\tau$ SFH leads to biased fits, often skewing the SFH peak to later cosmic times and thus underestimating the contribution from the older stellar populations.

To understand better the impact of the systematic biases introduced by the different SFH models, we compare three key global stellar properties ($M_\star$, sSFR, and $t_\mathrm{\star,\,mw}$) to the ground truth values from the TNG50 simulation. For the spatially resolved fits, we derived the global properties of a particular galaxy by summing or averaging across its pixel bins. We obtained $M_\star$ and SFR of a particular galaxy, by summing the values from each pixel bin, while $t_\mathrm{\star,\,mw}$ is computed as the stellar-mass-weighted average of the pixel ages. In Fig.~\ref{fig:fig_10}, we show the absolute differences in these properties relative to the TNG50 ground truth (top row), as well as the distribution of the absolute differences (bottom row) along with the mean offset ($\mu$), and scatter (standard deviation, $\sigma$). The values of $\mu$ and $\sigma$ for each global physical property are given in Table~\ref{tab:table_3}.

Focusing on the left column of Fig.~\ref{fig:fig_10}, we report that the global stellar masses inferred from the nonparametric model, at both resolved (SFH$_\mathrm{res,\,np}$) and integrated (SFH$_\mathrm{glob,\,np}$) scales, are closest to the ground truth, with a mean offset near zero. Conversely, the stellar masses inferred from the resolved parametric SFH$_\mathrm{res,\,\tau}$ and global parametric SFH$_\mathrm{glob,\,\tau}$ are systematically underestimated by 0.12 and 0.06~dex, respectively. The underestimation of the stellar mass at global scales is expected \citep[e.g.,][]{Jain_2024MNRAS.527.3291J, Mosleh_2025ApJ...983..181M}, largely due to the outshining effect, where the younger stellar populations ($<10$~Myr) dominate the observed SED and outshine the older stars ($>200$~Myr). The underestimation of $M_\star$ is more prominent in the case of the delayed $\tau$ SFH model. However, a more flexible parametric model, such as a double power-law SFH, could be a reliable option at integrated scales \citep[e.g.,][]{Harvey_2025arXiv250405244H}. 

Interestingly, we measure the strongest deviation in the resolved parametric model despite the expectation that spatial resolution should mitigate outshining. \citet{Iglesias_Navarro_2025arXiv250604336I} similarly found that applying a delayed-$\tau$ model in a pixel-by-pixel analysis of JWST data led to significant deviations in inferred galaxy properties, highlighting the limitations of simple parametric SFHs. Other studies have demonstrated that resolved parametric SFHs can yield accurate global $M_\star$ when more flexible functional forms are used. For instance, \citet{EP-Abdurrouf} employed a double power-law model at resolved scales along with physically motivated priors, resulting in very accurate global $M_\star$ for the same sample of 25 TNG50 galaxies. Similarly, \citet{Mosleh_2025ApJ...983..181M} reported that the double power-law SFH at resolved scales consistently outperformed other parametric SFH models, concluding that such parameterization is essential for accurate resolved SED fitting. In contrast, our analysis uses a simpler delayed-$\tau$ model, which limits accuracy and likely explains the larger deviations from the ground truth.

The middle column of Fig.~\ref{fig:fig_10} shows the comparison between the fitted sSFR and the ground truth TNG50 values. The present SFR is used here to calculate the sSFR. The sSFR values inferred from SFH$_\mathrm{res,\,np}$ are in excellent agreement with the TNG50 ground truth, with a mean offset of $\mu = 0.01$~dex. In contrast, the largest discrepancies arise in the case of SFH$_\mathrm{res,\,\tau}$, which systematically overestimates the sSFR. This overestimation is mainly driven by the combination of higher inferred SFRs (see Fig.~\ref{fig:fig_7}) and lower stellar masses returned by the delayed-$\tau$ model, both of which bias the sSFR upward. Furthermore, the comparison reveals that the sSFR estimates from SFH$_\mathrm{glob,\,np}$ and SFH$_\mathrm{glob,\,\tau}$ are systematically lower than the ground truth (the peak of the residual distribution is below 0~dex), yet likely for different reasons. In the case of SFH$_\mathrm{glob,\,np}$, we notice that the mass-weighted ages are on average older than the ground truth by 0.02~dex, resulting in lower SFRs and consequently lower sSFR (since $M_\star$ is properly recovered). Conversely, in the case of SFH$_\mathrm{glob,\,\tau}$ the systematically lower $M_\star$ and present SFR (see Fig.~\ref{fig:fig_7}) result in the underestimation of the sSFR. For quiescent galaxies ($\log_{10}(\mathrm{sSFR_{TNG50}/yr^{-1}}) \le -11$), SFH$_\mathrm{res,\,np}$ and SFH$_\mathrm{res,\,\tau}$ overestimate the sSFR, despite these galaxies having little to no active star formation. This inaccuracy may arise from the limited number of stellar particles in the latest TNG50 time bins, which can produce spurious SFR values at spatial scales, especially for quiescent galaxies. Uncertainties in the stellar ages of individual pixels, linked to the dust-age degeneracy, may further contribute to the overestimation of sSFR in galaxies with minimal star formation (see Fig.~\ref{fig:fig_5}). On the other hand, the sSFR recovered by both SFHs at global scales may suggest that for passive galaxies, simpler parametric models (e.g., such as $\tau$ models) can yield accurate integrated parameters.

Finally, the $t_\mathrm{\star,\,mw}$ estimates from SFH$_\mathrm{res,\,np}$ are the closest to the TNG50 ground truth, with a mean offset of $\mu = 0.01$~dex. The SFH$_\mathrm{glob,\,np}$ also performs well, with $\mu = 0.02$~dex. It is worth noting that in both nonparametric models, a potential correlation appears between the residuals and age, where younger ages tend to be overestimated, while older ages are underestimated within 0.1~dex. On the other hand, $t_\mathrm{\star,\,mw}$ estimates from SFH$_\mathrm{res,\,\tau}$ and SFH$_\mathrm{glob,\,\tau}$ show significant deviations, exceeding 0.2~dex. This discrepancy results from the skewed parametric SFHs toward later cosmic times, obscuring the extended SFHs of early-type galaxies and significantly affecting the inferred $t_\mathrm{\star,\,mw}$. These findings align with \citet{Suess_2022ApJ...935..146S}, \citet{Jain_2024MNRAS.527.3291J}, and \citet{Q1-SP031}, who also reported extremely young inferred stellar ages from SFHs derived on integrated scales with simple parametric SFH models.

\section{\label{sc:discussion} Discussion}

A key takeaway from this work is that nonparametric SFHs provide a more effective means of mitigating the outshining effect from the young stellar populations ($<10$~Myr), and therefore improving the accuracy of inferred galaxy properties. However, this improvement should not be overinterpreted, since our results are based on a small sample of galaxies. The primary advantage of spatially resolved SED fitting lies in its ability to recover the internal color variations from the different stellar populations and dust attenuation, capturing the complexity of galaxy evolution that is often missed in global analyses. Certainly, parametric SFH models on spatially resolved scales remain valuable, as they are less computationally intensive, and depending on the functional form can provide a robust reconstruction of SFHs, and yield less biased global galaxy properties \citep[e.g.,][]{Gimenez_Arteaga_2023ApJ...948..126G, Gimenez_Arteaga_2024A&A...686A..63G, Jain_2024MNRAS.527.3291J, Mosleh_2025ApJ...983..181M, EP-Abdurrouf}.

In this study, we show evidence that nonparametric modeling on resolved and global scales can recover the SFHs and derived properties with higher fidelity, making it a powerful tool for understanding galaxy formation and evolution. In some cases, SFH$_\mathrm{glob,\,np}$ provides a closer match to the TNG50 ground truth than SFH$_\mathrm{res,\,np}$. This effect likely arises from a combination of factors: residual photometric noise and de-blending uncertainties at the pixel level, stronger sensitivity to outshining in individual regions, and the fact that the integrated light tends to smooth out these localized mismatches. While the resolved fits allow us to recover spatial variations in SFHs, the global fits can sometimes provide a closer match to the average SFH shape, particularly for galaxies with low levels of ongoing star formation.

Our choice to focus on a subset of 25 galaxies from the TNG50-{\tt SKIRT} Atlas was motivated by the need to balance computational feasibility with sufficient diversity in morphology and star formation properties to test the pixel-by-pixel SED fitting framework. While this limited sample size prevents us from drawing definitive conclusions about the impact of spatial resolution on integrated quantities, it serves as a robust proof of concept for the methodology. The selected galaxies are biased toward relatively massive systems with near-solar metallicities and older stellar populations, reflecting both the stellar mass function of TNG50 at $z=0$ and the fact that galaxies with reliable radiative transfer post-processing typically require sufficient stellar particle sampling. Extending this work to include lower-mass, metal-poor galaxies would certainly broaden the dynamic range of stellar population properties, but such an expansion lies beyond the scope of the present study, which is intended as a methodological benchmark in preparation for \Euclid.

Furthermore, one persistent limitation of our approach is the dust-age degeneracy discussed in \ref{subsc:recovery_of_stellar_props}, which is partly driven by the use of a simplified, parametric attenuation law in the SED fitting, whereas {\tt SKIRT} models the full radiative transfer and dust geometry. This intentional mismatch reflects a realistic challenge faced in observational studies, where the true dust geometry is unknown and must be approximated by empirical or flexible attenuation models. Thus, the resulting degeneracies are not a flaw of the method but a realistic feature of the problem. A direct validation of the recovered dust attenuation was not feasible here because the current TNG50-{\tt SKIRT} Atlas only provides diffuse dust attenuation maps. The attenuation in star-forming regions is modeled separately, preventing a consistent reconstruction of the total attenuation. Updated radiative transfer simulations of the TNG50-{\tt SKIRT} Atlas are currently underway to address this limitation and include the dust emission in the far-infrared, enabling a more accurate assessment of dust attenuation in future work.

The broader implications of these results extend to testing models of galaxy evolution and evaluating the internal consistency of observational datasets. For instance, as highlighted by \citet{Fu_2024MNRAS.532..177F}, in a $\Lambda$CDM paradigm of structure formation, galaxies evolve hierarchically within dark matter halos via mergers and star formation. This growth is reflected in a set of interconnected observables, most notably the stellar mass functions, the SFHs, the satellite galaxy abundances, the merger rates, and even other such properties as intra-cluster light, each one capturing different aspects of the galaxy formation process. A physically sound model of galaxy evolution must be able to simultaneously predict and reproduce all these observables in a consistent way. However, this challenge is often complicated by inconsistencies among datasets, driven by differences in observational methods and assumptions. As demonstrated here, \Euclid will offer the opportunity to measure nonparametric SFHs, stellar mass functions, as well as intra-cluster light from the same datasets across large areas and at different cosmic epochs, allowing for 1) testing the self-consistency of these data and 2) validating hierarchical models of galaxy formation. The rich and varied datasets from \Euclid, including the SFHs put forward here, will help identify systematic tensions among differential and integrated quantities (SFHs versus stellar mass functions, for example), guiding the refinements in both modeling and data calibration. Ultimately, such holistic approaches will provide a critical bridge between observation and theory, enabling the development of galaxy evolution models that are not only descriptive but also predictive, robust, and self-consistent across cosmic time.

\section{\label{sc:sum_conclusions} Summary and conclusions}

\Euclid will observe approximately 14\,000~$\mathrm{deg}^{2}$ of the extragalactic sky, delivering high-resolution imaging that enables spatially resolved studies of nearby galaxies. To assess the feasibility of reconstructing accurate SFHs from \Euclid-like data, we presented detailed measurements of SFHs on both global and spatially resolved scales for a sample of 25 TNG50-{\tt SKIRT} Atlas galaxies using {\tt Prospector}. We examine the SFHs derived via (1) resolved SED fitting using a nonparametric SFH model (SFH$_\mathrm{res,\,np}$), (2) resolved SED fitting using a commonly adopted parametric $\tau$-model (SFH$_\mathrm{res,\,\tau}$), (3) global SED fitting with a nonparametric model (SFH$_\mathrm{glob,\,np}$), and (4) global SED fitting using a parametric SFH (SFH$_\mathrm{glob,\,\tau}$). We applied these models to mock UV-NIR photometry derived from the TNG50 cosmological simulation and processed with the radiative transfer code {\tt SKIRT}. This approach provides a proof of concept for extracting spatially resolved SFHs of local galaxies with \Euclid, highlighting the strengths and limitations of nonparametric SFH modeling in the context of next-generation galaxy surveys. Our main findings and conclusions are the following:

\begin{itemize}

\item The mass surface density is recovered with high accuracy using broadband photometry, thanks to the NIR coverage from \Euclid (mean offset $\mu=0.01$~dex and scatter $\sigma=0.1$~dex from the ground truth values). The SFR surface density and stellar ages are reasonably recovered ($\mu\le0.04$~dex), considering the fact that we only fit broadband photometry without using an age-mass prior. A secondary peak in the stellar age distribution is observed, where {\tt Prospector} assigns more dust to explain the redder colors in the spatial bins. Finally, we show that photometry-only fits with a linear $Z_\star$ prior function do not yield meaningful constraints on $Z_\star$ for individual galaxies, although they effectively avoid a downward bias. While the precision remains limited, the accuracy is generally good. We also note that the results for $Z_\star$ are highly sensitive to the choice of prior function. 
\\
\item In Figs.~\ref{fig:fig_7} and \ref{fig:fig_8}, we find that the best method to reconstruct the SFHs of nearby galaxies, when using photometry alone, is employing a nonparametric SFH model at spatially resolved scales (SFH$_\mathrm{res,\,np}$). Indeed, the nonparametric SFH, at resolved scales, is in very good agreement with the ground truth SFH from TNG50. The nonparametric SFH at global scales appears to be a reliable method as well, returning robust stellar mass and age estimates. The flexibility of the nonparametric model allows it to better capture the different variations and amplitudes of star formation across cosmic time, while being less affected by outshining from recent star formation activity. 
\\
\item Based on the formation timescales derived from the different SFH models, we find that nonparametric SFHs at global scales (SFH$_\mathrm{glob,\,np}$) are generally more reliable for reconstructing SFHs when spatial resolution is not available. In contrast, the delayed $\tau$-model, at both resolved (SFH$_\mathrm{res,\,\tau}$) and integrated (SFH$_\mathrm{glob,\,\tau}$) scales, tends to fail in capturing early star formation activity, leading to significant deviations in metrics such as $t_{20}$ and $t_{50}$ compared to the ground truth SFHs from TNG50 (see Table~\ref{tab:table_3}). This limitation results in systematic underestimation of both the stellar mass and the average age of the stellar populations. Nonetheless, the $\tau$-model remains useful for characterizing recent star formation activity, as UV-bright stars dominate the global SED.
\\
\item We quantified the systematics in three key global properties (i.e., stellar masses, sSFR, and stellar ages) as derived from the different SFH treatments, by comparing them to the ground truth values from TNG50 (see Table~\ref{tab:table_3}). Overall, we find that both stellar mass and stellar age are severely underestimated compared to the ground truth values when the simple $\tau$-model is used, while deviations are reduced when using the nonparametric SFH model, especially at resolved scales. In particular, for the global stellar mass, we measure a mean offset of $\mu=-0.03$, $\mu=-0.12$, $\mu=0.01$, and $\mu=-0.06$ for the SFH$_\mathrm{res,\,np}$, SFH$_\mathrm{res,\,\tau}$, SFH$_\mathrm{glob,\,np}$, and SFH$_\mathrm{glob,\,\tau}$, respectively. For the sSFR, we find deviations of $\mu=0.01$, $\mu=0.4$, $\mu=-0.17$, and $\mu=-0.24$ for the SFH$_\mathrm{res,\,np}$, SFH$_\mathrm{res,\,\tau}$, SFH$_\mathrm{glob,\,np}$, and SFH$_\mathrm{glob,\,\tau}$, respectively. For quiescent galaxies, the difference in sSFR recovered by SFH$_\mathrm{glob,\,\tau}$ is reduced, suggesting that simpler parametric models can yield accurate integrated parameters. Finally, for the mass-weighted ages, we estimate a mean offset of $\mu=0.01$, $\mu=-0.27$, $\mu=0.02$, and $\mu=-0.23$ for the SFH$_\mathrm{res,\,np}$, SFH$_\mathrm{res,\,\tau}$, SFH$_\mathrm{glob,\,np}$, and SFH$_\mathrm{glob,\,\tau}$, respectively.

\end{itemize}

In summary, spatially resolved nonparametric SFHs provide a more effective approach to mitigating outshining effects, offering improved accuracy in the determination of galaxy stellar properties. This highlights the importance of using spatially resolved analyses to fully capture the complex evolutionary pathways of galaxies and avoid biases inherent in global parametric models. In the context of the \Euclid mission, this study demonstrates that photometry-only fits, when combined with nonparametric models and spatial resolution, can enable realistic reconstructions of galaxy SFHs, paving the way for transformative insights into galaxy formation and evolution. Given the complexity and computational demands of spatially resolved SED fitting with a nonparametric SFH, our method is focused on the quality of the inferred galactic properties and can be applied to a small, yet still significant, number of galaxies in the local Universe using deep, high-resolution imaging data from \Euclid, complemented by optical observations from other surveys such as UNIONS, DES, and LSST, as well as UV observations from GALEX. The wide sky coverage provided by these surveys will enable the analysis of a diverse range of galaxies, residing in different environments and encompassing a broad range of stellar properties and morphologies.

\begin{acknowledgements}
  
We thank the anonymous referee for the valuable remarks and suggestions that helped us improve the paper. AN gratefully acknowledges the support of the Belgian Federal Science Policy Office (BELSPO) for the provision of financial support in the framework of the PRODEX Programme of the European Space Agency (ESA) under contract number 4000143347. EDC acknowledges funding from the European Union grant WIDERA ExGal-Twin, GA 101158446.
\AckDatalabs
\AckEC \AckCosmoHub

\end{acknowledgements}

%
%

\bibliography{Euclid}

%

  

%

\begin{appendix}
%

\onecolumn

\section{\label{apdx:A} The effect of prior functions on stellar metallicities}

\begin{figure}[h!]
    \centering
    \includegraphics[width=\textwidth]{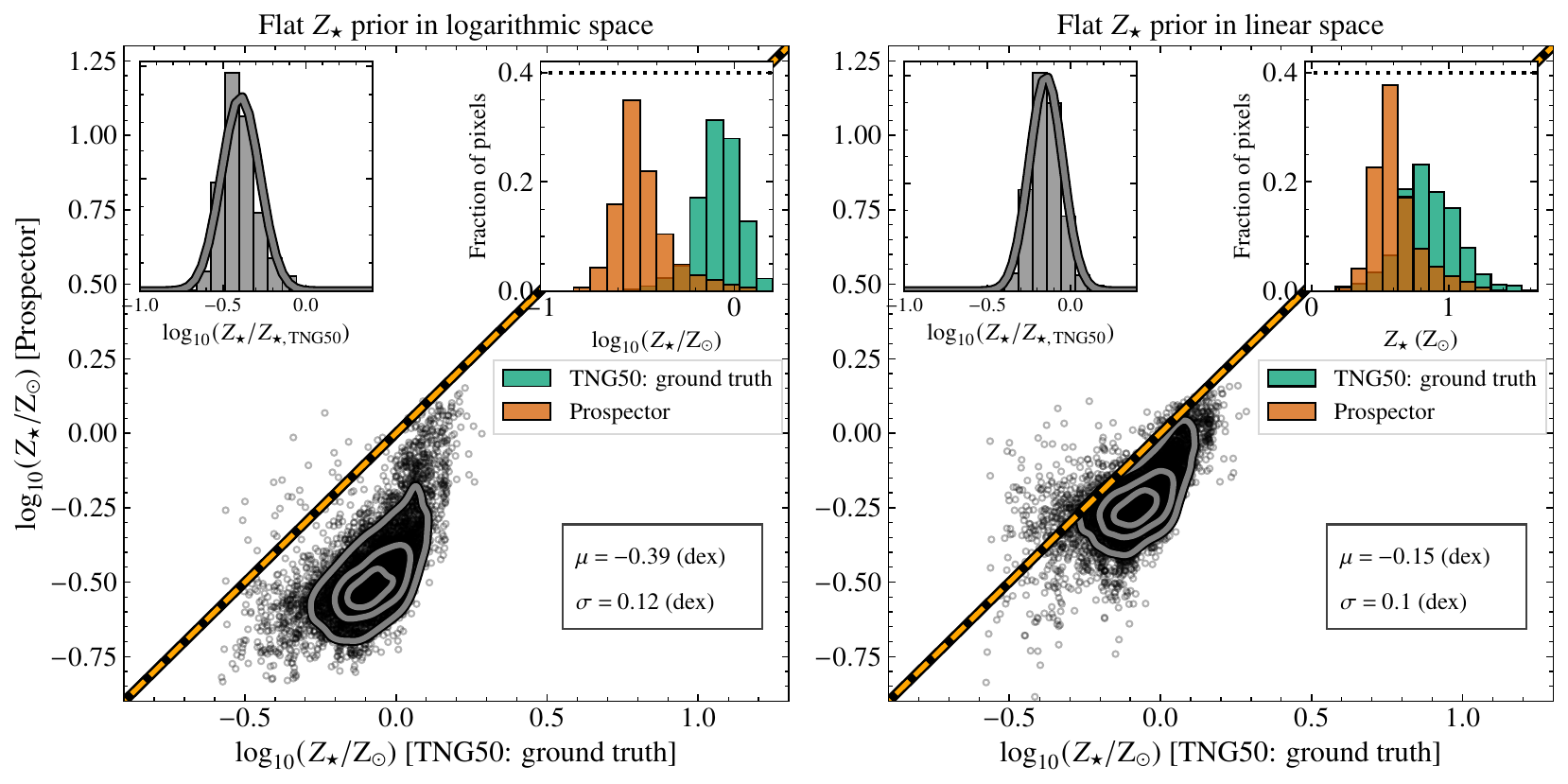}
    \caption{Comparison of the spatially resolved stellar metallicities for 6958 pixel bins in the 25 sample galaxies measured by different prior functions. The true values of $Z_\star$ ($x$-axis) are compared with those from our fiducial run ($y$-axis), when a flat prior for $Z_\star$ is used in logarithmic space (left panel) and in linear space (right panel). The dashed orange line shows the one-to-one relation, while contours enclose 20\%, 50\%, and 80\% of the total data. Top-left: Residual distribution, $\logten \left(Z_\mathrm{\star}/Z_\mathrm{\star,\, TNG50}\right)$. Top-right: Comparison of the true $Z_\star$ values (teal distribution) with those recovered from {\tt Prospector} (orange distribution). The dotted lines correspond to the prior probability distributions, scaled with an arbitrary factor. The mean offset ($\mu$) and scatter ($\sigma$) between the two distributions are indicated in each panel. A flat prior in linear space for $Z_\star$ returns values closer to the ground truth.}
    \label{fig:fig_a1}
\end{figure}

In this section, we test the effect of the prior function on the stellar metallicity estimates. In a recent study, \citet{Nersesian_2025A&A...695A..86N} showed that when fitting photometry alone, a flat metallicity prior in linear space better recovers the distribution of metallicities measured from the spectrophotometric analysis of the LEGA-C survey \citep{van_der_Wel_2021ApJS..256...44V}. In fact when fitting spectroscopy alone or together with photometry, the results are unaffected by the choice of the $Z_\star$ prior (i.e., whether in linear or logarithmic space). In Fig.~\ref{fig:fig_a1}, we demonstrate that using a flat prior in linear space for $Z_\star$ results in measured metallicities that better align with those from our true values from TNG50. For this comparison, all stellar metallicities have been rescaled to the Solar metallicity value of $\mathrm{Z}_\odot = 0.02$. When a flat prior in linear space is used for $Z_\star$, the mean offset ($\mu$) from the fiducial $Z_\star$ is considerably reduced, from 0.39~dex to 0.15~dex. The scatter ($\sigma$) around the mean remains similar yet slightly reduced by 0.02~dex. This improvement in the $Z_\star$ measurements does not necessarily imply a better constraint on the stellar metallicities. Nevertheless, \citet{Nersesian_2024A&A...681A..94N} showed that stellar metallicities are difficult to be recovered from multiwavelength photometry, even when there is information from both broadbands and narrowbands.

\section{\label{apdx:B} The benefit of \Euclid NIR photometry} 

\begin{figure*}[t]\label{LastPage}
    \centering
    \includegraphics[width=\textwidth]{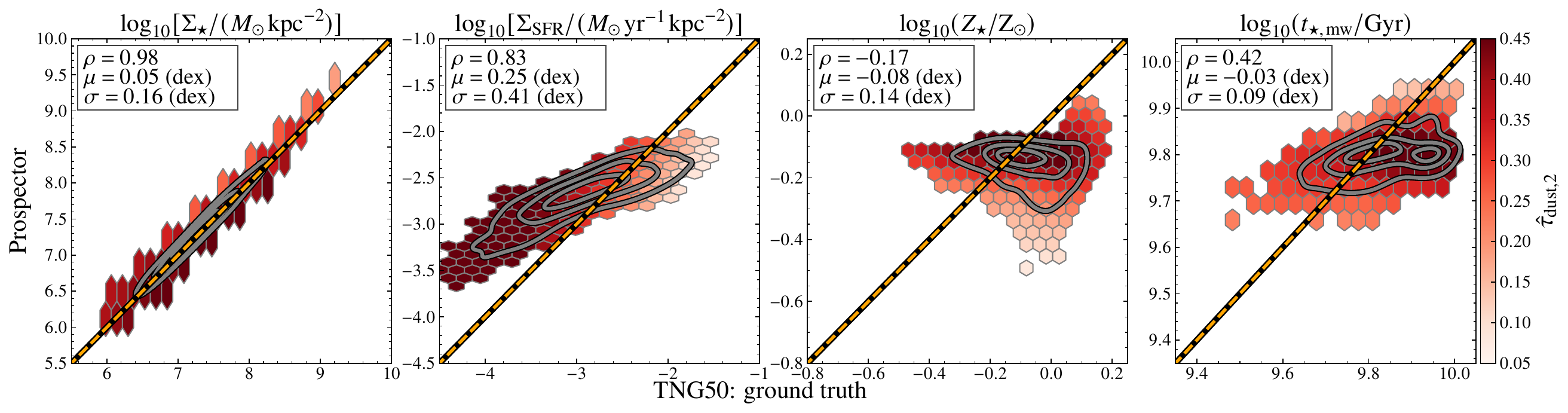}
    \caption{Comparisons of the spatially resolved stellar population properties derived from spatially resolved SED fitting with a nonparametric SFH model on seven broadbands. The photometric bands used in this analysis include the GALEX FUV and NUV, and LSST $u$, $g$, $r$, $i$, and $z$, in the resolution of GALEX. From left to right, we show the pixel values of the stellar property maps, that is $\Sigma_\star$, $\Sigma_\mathrm{SFR}$, $Z_\star$, and $t_\mathrm{\star,\,mw}$. Points are color-coded according to the average trend of the dust optical depth in the $V$ band ($\hat{\tau}_\mathrm{dust,\,2}$), calculated with the {\tt LOESS} method. Contours enclose 20\%, 50\%, and 80\% of the total data. The dashed orange line shows the one-to-one relation. The Spearman’s correlation coefficient ($\rho$), measured bias (average offset $\mu$), and scatter (standard deviation $\sigma$) relative to the ground truth from TNG50 are indicated in the legend of each panel.}
    \label{fig:fig_b1}
\end{figure*}


Similar to Fig.~\ref{fig:fig_5}, Fig.~\ref{fig:fig_b1} presents the distribution of all spatial pixel bins across the 25 galaxies in our sample, but this time derived from SED fitting performed without the \Euclid photometric bands. The mean offset ($\mu$) and scatter (standard deviation, $\sigma$) of the logarithmic ratio between the best-fit parameters and the ground truth, along with the Spearman rank-order correlation coefficient ($\rho$) for each stellar property, are summarized in the top-left corner of each panel in Fig.~\ref{fig:fig_b1}. Points are color-coded according to the average trend of the dust optical depth in the $V$ band ($\hat{\tau}_\mathrm{dust,\, 2}$), calculated with the {\tt LOESS} method.

Excluding the \Euclid bands introduces a systematic offset of 0.05~dex in $\Sigma_\star$ relative to the ground truth, and an increased scatter around the mean ($\sigma = 0.16$~dex), indicating that the \Euclid bands play a crucial role in obtaining reliable stellar mass estimates. For $\Sigma_\mathrm{SFR}$, the recovery worsens significantly, with a larger mean offset ($\mu = 0.25$~dex) and increased scatter ($\sigma = 0.41$~dex). As seen in Fig.~\ref{fig:fig_5}, the lower end of the $\Sigma_\mathrm{SFR}$ distribution tends to be overestimated, but this effect is even more pronounced in the absence of the \Euclid bands. Additionally, we observe a stronger correlation between $\Sigma_\mathrm{SFR}$ and $\hat{\tau}_\mathrm{dust,\, 2}$, suggesting that the dust-age degeneracy becomes more severe when NIR information is missing.

Furthermore, the exclusion of \Euclid photometry results in a completely unconstrained $Z_\star$, whereas the recovery of stellar ages remains largely unaffected. We measure an offset ($\mu = -0.03$~dex) and scatter ($\sigma = 0.09$~dex), consistent with the results in Fig.~\ref{fig:fig_5}. However, we again observe the formation of a secondary peak in the stellar age distribution, which is linked to the dust-age degeneracy. In these cases, {\tt Prospector} infers younger ages, attributing the red colors primarily to dust attenuation rather than the presence of older stellar populations.
\end{appendix}
\end{document}